\documentclass[letter]{aa}  
\usepackage{graphicx}
\usepackage{txfonts}
\usepackage{lipsum}
\usepackage{subcaption}
\usepackage{lscape}
\usepackage{placeins}

\begin{document}
\nolinenumbers
\makeatletter
\let\linenumbers\relax
\let\linenumberfont\relax
\makeatother

\DeclareRobustCommand{\orcidicon}{%
    \begin{tikzpicture}
    \draw[lime, fill=lime] (0,0) 
    circle [radius=0.16] 
    node[white] {{\fontfamily{qag}\selectfont \tiny ID}};
    \draw[white, fill=white] (-0.0625,0.095) 
    circle [radius=0.007];
    \end{tikzpicture}
    \hspace{-2mm}
}

\newcommand*{\tabhead}[1]{\multicolumn{1}{c}{\bfseries #1}}

\newcommand{\mystar}{{\Large {\fontfamily{lmr}\selectfont$\star$}}}
\newcommand{\pds}{PDS~70}

\newcommand{\jgrp}{JGR-Planets}
\newcommand{\rsos}{Royal Soc. Open Sci.}
\newcommand{\astrolett}{Astron. Lett.}
\newcommand{\psj}{Planet. Sci. J.}
\newcommand{\jatis}{J. Astron. Telesc. Instrum. Syst.}

\newcommand{\mgtwo}{Mg\,II}
\newcommand{\ctwo}{C\,II}
\newcommand{\molh}{H$_2$}
\newcommand{\lyalph}{Ly-$\alpha$}
\newcommand{\fetwo}{Fe\,II}

\newcommand{\lognh}{$\log(N_{\rm H~I})$}
\newcommand{\oi}{[O\,I]}
\newcommand{\suii}{[S\,II]}
\newcommand{\nii}{[N\,II]}
\newcommand{\nai}{Na\,I}
\newcommand{\caii}{Ca\,II}
\newcommand{\mdotacc}{$\cdot{M}_{\rm acc}$}
\newcommand{\mdotwind}{$\cdot{M}_{\rm wind}$}
\newcommand{\hei}{He\,I}
\newcommand{\lyalpha}{Ly-$\alpha$}
\newcommand{\halpha}{H$\alpha$}
\newcommand{\mgi}{Mg\,I}
\newcommand{\sii}{Si\,I}
\newcommand{\water}{H$_2$O}
\newcommand{\methane}{CH$_4$}
\newcommand{\cotwo}{CO$_2$}
\newcommand{\htwo}{H$_2$}


\newcommand{\rosat}{\emph{Rosat}}
\newcommand{\galex}{\emph{GALEX}}
\newcommand{\tess}{\emph{TESS}}
\newcommand{\plato}{\emph{PLATO}}
\newcommand{\gaia}{\emph{Gaia}}
\newcommand{\ktwo}{\emph{K2}}
\newcommand{\jwst}{\emph{JWST}}
\newcommand{\kepler}{\emph{Kepler}}
\newcommand{\corot}{\emph{CoRoT}}
\newcommand{\hipp}{\emph{Hipparcos}}
\newcommand{\spitzer}{\emph{Spizter}}
\newcommand{\herschel}{\emph{Herschel}}
\newcommand{\hst}{\emph{HST}}
\newcommand{\wise}{\emph{WISE}}
\newcommand{\swift}{\emph{Swift}}
\newcommand{\chandra}{\emph{Chandra}}
\newcommand{\xmm}{\emph{XMM-Newton}}

\newcommand{\twa}{TW Hydra}
\newcommand{\bpic}{$\beta$~Pictoris}
\newcommand{\abdor}{AB~Doradus}
\newcommand{\rup}{Ruprecht~147}
\newcommand{\etacha}{$\eta$\,Chamaeleontis}
\newcommand{\usco}{Upper~Sco}
\newcommand{\rhooph}{$\rho$~Oph}

\newcommand{\doar}{DoAr\,25}
\newcommand{\epcha}{EP\,Cha}
\newcommand{\rylup}{RY\,Lup}
\newcommand{\hdtwofour}{HD\,240779}

\newcommand{\sigrv}{$\sigma_{\rm RV}$}

\newcommand{\msunyr}{\rm{M_{\sun} \, yr^{-1}}}
\newcommand{\etal}{\mbox{\rm et al.~}}
\newcommand{\ms}{\mbox{m\,s$^{-1}~$}}
\newcommand{\kms}{\mbox{km\,s$^{-1}~$}}
\newcommand{\ks}{\mbox{km\,s$^{-1}~$}}
\newcommand{\kse}{\mbox{km\,s$^{-1}$}}
\newcommand{\mse}{\mbox{m\,s$^{-1}$}}
\newcommand{\msy}{\mbox{m\,s$^{-1}$\,yr$^{-1}~$}}
\newcommand{\msye}{\mbox{m\,s$^{-1}$\,yr$^{-1}$}}
\newcommand{\msun}{M$_{\odot}$}
\newcommand{\msune}{M$_{\odot}$}
\newcommand{\rsun}{R$_{\odot}$}
\newcommand{\lsun}{L$_{\odot}~$}
\newcommand{\rsune}{R$_{\odot}$}
\newcommand{\mjup}{M$_{\rm JUP}~$}
\newcommand{\mjupe}{M$_{\rm JUP}$}
\newcommand{\msat}{M$_{\rm SAT}~$}
\newcommand{\msate}{M$_{\rm SAT}$}
\newcommand{\mnep}{M$_{\rm NEP}~$}
\newcommand{\mnepe}{M$_{\rm NEP}$}
\newcommand{\mearth}{$M_{\oplus}$}
\newcommand{\mearthe}{$M_{\oplus}$}
\newcommand{\rearth}{$R_{\oplus}$}
\newcommand{\rearthe}{$R_{\oplus}$}
\newcommand{\rjup}{R$_{\rm JUP}~$}
\newcommand{\msinie}{$M_{\rm p} \sin i$}
\newcommand{\vsinie}{$V \sin i$}
\newcommand{\mbsini}{$M_b \sin i~$}
\newcommand{\mcsini}{$M_c \sin i~$}
\newcommand{\mdsini}{$M_d \sin i~$}
\newcommand{\chisq}{$\chi_{\nu}^2$}
\newcommand{\chinu}{$\chi_{\nu}$}
\newcommand{\chinusq}{$\chi_{\nu}^2$}
\newcommand{\arel}{$a_{\rm rel}$}
\newcommand{\feh}{\ensuremath{[\mbox{Fe}/\mbox{H}]}}
\newcommand{\rphk}{\ensuremath{R'_{\mbox{\scriptsize HK}}}}
\newcommand{\lrphk}{\ensuremath{\log{\rphk}}}
\newcommand{\cs}{$\sqrt{\chi^2_{\nu}}$}
\newcommand{\etaearth}{$\mathbf \eta_{\oplus} ~$}
\newcommand{\etaearthe}{$\mathbf \eta_{\oplus}$}
\newcommand{\searth}{$S_{\bigoplus}$}
\newcommand{\mdotyr}{$M_{\odot}$~yr$^{-1}$}
\newcommand{\micron}{$\mu$m}

\newcommand{\sini}{\ensuremath{\sin i}}
\newcommand{\msini}{\ensuremath{M_{\rm p} \sin i}}
\newcommand{\mplsini}{\ensuremath{\mpl\sin i}}
\newcommand{\teff}{\ensuremath{T_{\rm eff}}}
\newcommand{\teq}{\ensuremath{T_{\rm eq}}}
\newcommand{\logg}{\ensuremath{\log{g}}}
\newcommand{\vsini}{\ensuremath{v \sin{i}}}
\newcommand{\ebv}{E($B$-$V$)}

\newcommand{\Kepler}{\emph{Kepler}~}
\newcommand{\Keplere}{\emph{Kepler}}
\newcommand{\blender}{{\tt BLENDER}~}

\newcommand{\kp}{\ensuremath{\mathrm{Kp}}}
\newcommand{\rstar}{\ensuremath{R_\star}}
\newcommand{\mstar}{\ensuremath{M_\star}}
\newcommand{\loggstar}{\ensuremath{\logg_\star}}
\newcommand{\lstar}{\ensuremath{L_\star}}
\newcommand{\astar}{\ensuremath{a_\star}}
\newcommand{\loglstar}{\ensuremath{\log{L_\star}}}
\newcommand{\rhostar}{\ensuremath{\rho_\star}}

\newcommand{\rp}{\ensuremath{R_{\rm p}}}
\newcommand{\rpl}{\ensuremath{r_{\rm P}}~}
\newcommand{\rple}{\ensuremath{r_{\rm P}}}
\newcommand{\mpl}{\ensuremath{m_{\rm P}}~}
\newcommand{\lpl}{\ensuremath{L_{\rm P}}~}
\newcommand{\rhopl}{\ensuremath{\rho_{\rm P}}~}
\newcommand{\loggpl}{\ensuremath{\logg_{\rm P}}~}
\newcommand{\logrpl}{\ensuremath{\log r_{\rm P}}~}
\newcommand{\logrple}{\ensuremath{\log r_{\rm P}}}
\newcommand{\loga}{\ensuremath{\log a}~}
\newcommand{\logmpl}{\ensuremath{\log m_{\rm P}}~}

\newcommand{\fludensunits}{ergs s$^{-1}$ cm$^{-2}$ \AA$^{-1}$}

\title{Water-Cooled (sub)-Neptunes Get Better Gas Mileage}
\author{
Tatsuya Yoshida\inst{\ref{tohuku}}
\and 
Eric Gaidos\inst{\ref{hawaii},\ref{vienna}}
}
\institute{
Faculty of Science, Tohoku University, Sendai, Miyagi 980-8578, Japan\label{tohuku}\\
\email{tatsuya@tohoku.ac.jp}
\and
Department of Earth Sciences, University of Hawai'i at M\"{a}noa, Honolulu, Hawai'i 96822 USA\label{hawaii}
\email{gaidos@hawaii.edu}
\and
Institute for Astrophysics, University of Vienna, 1180 Vienna, Austria\label{vienna}\\
}
\date{Received September 30, 20XX}
 
\abstract
{The demographics of sub-Jovian planets around low-mass stars is dominated by populations of ``sub-Neptunes" and ``super-Earths", distinguished by the presence or absence of envelopes of low-molecular weight volatiles, i.e., \htwo, He, and \water.  The current paradigm is that sub-Neptunes on close-in orbits evolve into super-Earths via atmospheric escape driven by high-energy stellar irradiation.   We use an integrated hydrodynamic-radiation-chemical network model of outflow to demonstrate that this escape is modulated by the abundance of \water, an efficient infrared coolant.  Increasing \water/\htwo\ at the base of the flow induces an order-of-magnitude decline in escape rate, with definitive consequences for retention of envelopes over Gyr.  We show that saturation limits on \water\ in the upper atmospheres of temperate sub-Neptunes could explain the paradoxical observations that these objects disappear more rapidly than their counterparts closer to their host stars.   We also propose that the scarcity of sub-Neptunes around very low mass stars could be related to the water-poor chemistry of their antecedent protoplanetary disks.  Observations of atmospheric \water\ by \jwst\ as well as searches for atmospheric escape from younger planets using H and He lines could test these predictions.}
\keywords{Planets and satellites: atmospheres ---  Planets and satellites: physical evolution --- Planet-star interactions}
\maketitle

\section{Introduction}
\label{sec:intro}
A photometric survey for transiting exoplanets by  \kepler\ showed that Earth- to Neptune-size planets are far more numerous on close-in orbits around solar-type stars than Jovian counterparts \citep{Howard2012,Petigura2013}.  Spectroscopic refinement of host star (and hence transiting planet) radii confirmed that this population consists of distinct ``super-Earth" and ``sub-Neptune" moieties with modal radii of about 1.3 and 2.2\rearth\ separated by a sparsely-populated valley at $\sim$1.7\rearth\ \citep{Owen2013,Fulton2017,Petigura2022}.  This distribution appears to be universal among low-mass stars including M dwarfs \citep{Cloutier2020,Hirano2021,Venturini2024,Gaidos2024c}.

Planets much larger than the Earth but smaller than Neptune have no Solar System analog and their formation, composition, and evolution are active areas of investigation in exoplanet research.    Available combinations of both mass and radius measurements indicate that super-Earths are primarily rocky while sub-Neptunes have thick, low-density envelopes comprised of low molecular weight molecules, particularly \htwo/He or \water, around a rocky/icy core \citep{Weiss2014}.  In the current paradigm, sub-Neptunes can lose their envelopes and evolve into super-Earths, and there is at least tentative evidence for atmospheric escape from some objects \citep[e.g.,][]{Gaidos2022a,Zhang2023,Rockcliffe2023,Orell-Miquel2023,Masson2024}.  

Photoevaporation of a planetary atmosphere occurs when its upper layers are heated by stellar extreme ultraviolet (EUV) radiation\footnote{X-rays and far-UV radiation make only minor contributions to heating \citep[e.g.,][]{Owen+2020}.} to the energies required for escape.  Under sufficiently high irradiation, a collisional hydrodynamic flow can develop.  The temperature profile in such a flow is governed by a balance between absorption of stellar EUV radiation vs. work done by the outflow, plus radiative cooling by atmospheric components.  Radiative cooling is usually inefficient because absorption of X-rays and EUV invariably occurs above the homopause (or turbopause) where eddy diffusion can transport molecules against gravitational segregation.  On Earth, the most important coolant molecules are \cotwo\ and, more variably, nitric oxide \citep[NO,][]{Kockarts1980}.  High levels of \cotwo\ in the Archaean terrestrial atmosphere would have caused dramatically reduced upper atmosphere temperatures relative to the present, suppressing escape \citep{Johnstone2018}.   Recent anthropogenic increase in \cotwo\ has caused a small but measurable decline in mesosphere/lower thermosphere temperatures \citep{Roble1989,Mlynczak2022}.

\water\ has a lower molecular weight and is a more effective IR radiator than either \cotwo\ or NO, but on Earth, with an equilibrium temperature $T_e \approx 255$\,K, \water\ is trapped in oceans and by condensation at altitudes $\lesssim$15\,km, far below the homopause at 90 km.  On planets with higher equilibrium temperatures, especially those in a ``runaway" greenhouse state, \water\ is entirely in the atmosphere \citep{Kasting1988} and the Clausius-Clapeyron relation permits higher \water\ in the upper atmosphere.  In sub-Neptunes with envelopes of \htwo-He, super-runaway irradiation means that interior \water\ is in a supercritical phase that could be miscible with \htwo, allowing it to mix throughout the convective atmosphere \citep{Pierrehumber2023}.   Planets with inflated, steam-dominated atmospheres could be present in or even dominate the small planet population on close-in orbits \citep{Turbet2020,Piaulet-Ghorayeb2024}.  On the other hand, some planets may have accreted from oxygen-poor, dehydrated material and may lack \htwo\ altogether \citep{Gaidos2000,Raymond2007}.  

The effect of variation in \water\ abundance on atmospheric escape rates and the evolution of small planets, particularly sub-Neptunes, has not yet been investigated.   These effects could be observed around M dwarfs, where the orbital distance corresponding to water condensation falls within the limits of planet detection surveys.  Here, we use a model of EUV-driven outflow to demonstrate this effect for temperate sub-Neptunes around M dwarfs, and investigate how it can explain some aspects of exoplanet demographics. 

\section{Model}
\label{sec:model}
We used a radially symmetric 1-d model of hydrodynamic escape \citep{Yoshida+2020,Yoshida+2022}, modified for application to sub-Neptunes on close-in orbits around M dwarf stars.   Details of the model can be found in Appendix D and these papers. Here we outline the model, focusing on the modifications.  We calculated atmospheric escape rates on a coarse, 4-d grid with respect to the stellar EUV flux, planetary gravitational acceleration $g$, \water/\htwo\ ratio at the base of the flow, and planetary mass.  Using these rates, we evolved sub-Neptunes with a fixed planetary mass $M_p$ and on circular orbits around an M2-type (0.45\msun) star, accounting for the (pre-main sequence) evolution in stellar bolometric luminosity, and evolution in stellar EUV luminosity due to rotational spin-down.  We adopted $M_p$ of 2.5 and 5\mearth, initial envelope mass fractions of 1--5\%, and semi-major axes of 0.01--0.3\,au, corresponding to main-sequence stellar irradiance of 0.3--250 times the current terrestrial value. The effect of fixing $M_p$ on evolutionary calculations was minimal since the atmospheric mass fraction is small. The planetary radius $R_p$ was calculated as a function of envelope mass fraction, age, and stellar irradiance using the models of \citet{Lopez+2014}.  At each time-step, H and O escape rates were calculated by interpolation on the grid, and the envelope mass fraction was adjusted accordingly.  For atmospheric evolution calculations, we fixed the basal \water/\htwo\ ratio, but we investigated the effect of allowing it to increase as a result of H escape in a single, fiducial case.

The atmosphere below the base of the outflow was assumed to be composed only of H$_2$ and H$_2$O. Helium and other carbon- and nitrogen-bearing species are neglected. The lower and upper boundaries of the model are set at radial distances of $R_{p}$ and $50R_{p}$, respectively.   The number density of H$_2$ at the lower boundary was set at $1\times 10^{13}\,\mathrm{cm^{-3}}$, supposing the typical gas density at the altitude where stellar EUV irradiation is fully absorbed.   The H$_2$O/H$_2$ ratio was fixed between 0 and 0.1 in each simulation.  This is not equal to the \water/\htwo\ ratio in the deep atmosphere, i.e., at or below the turbopause where eddy diffusivity mixes atmosphere constituents, but is instead a lower limit (see Discussion).  The temperature at the lower boundary was set to the atmospheric skin temperature, which is the asymptotic temperature at high altitudes of the upper atmosphere that is optically thin in the thermal-IR and transparent to shortwave radiation \citep{Catling+2017}. This basal temperature is 400\,K at $\sim 10\times$ the current terrestrial irradiance. Although this temperature depends on semi-major axis, its variation minimally impacts the outflow structure and escape rate when the escape parameter, the ratio of the planetary gravitational potential energy to the thermal energy, exceeds $\sim$10 \citep{Kubyshkina+2018}, as is the case here.

The governing equations in the model are the fluid equations of continuity, momentum, and energy for a multi-component gas, plus those describing chemical reactions and radiative transport, assuming a spherically symmetric geometry. These equations were solved through time-dependent numerical integration until the physical quantities converge to steady profiles, following the calculation method in \citet{Yoshida+2020}.  The chemical species and reactions considered in this study were the same as in \citet{Yoshida+2022}. 93 chemical reactions involving 15 species (H$_2$, H$_2$O, H, O, O($^1$D), OH, O$_2$, H$^+$, H$_{2}^{+}$, H$_{3}^{+}$, O$^{+}$, O$_{2}^{+}$, OH$^{+}$, H$_{2}$O$^{+}$, and H$_3$O$^{+}$, Table \ref{tab:A1} and \ref{tab:A2}) are included.

Heating by absorption of stellar X-rays and UV radiation was calculated using a reconstruction of the 0.1--280\,nm spectrum of GJ\,832 as a representative M2 dwarf \citep{Loyd+2016}. The EUV flux at 1\,au is $1.1\times 10^{-3}\,\mathrm{W\,m^{-2}}$, about 25\% that of the Sun. For stellar EUV luminosity evolution, we adopt the model of \citet{Johnstone+2021} for a star with a mass of $0.4M_{\odot}$ and an initial rotation rate ten times that of the current solar value.  To represent both the higher EUV luminosity typical of young stars and a range of semi-major axes, we scaled this spectrum up to $10^4$ times the current terrestrial irradiance.  Heating and photolysis rate profiles were calculated by numerically solving the radiative transfer of parallel stellar photon beams in a spherically symmetric atmosphere, following the method of \citet{Tian+2005}. The resulting 3-d heating distribution was averaged over spherical shells. Stellar visible and IR radiation was neglected as a heat source in the upper atmosphere.  We included radiative cooling by thermal line emission from H$_2$O, OH, H$_{3}^{+}$, OH$^{+}$, and H$_3$O$^{+}$, using line data from the HITRAN \citep{Rothman+13} and ExoMol \citep{Tennyson+2016} databases. In addition to the line emission considered by \citet{Yoshida+2022}, we included thermal line emission from H$_3$O$^+$ and have expanded the line data for H$_2$O and OH to cover a broader temperature range (Appendix D). We calculated radiative cooling rates for these species using the method formulated by \citet{Yoshida+2020}. 

\section{Results}
\label{sec:results}
\subsection{Effect of \water\ abundance on atmospheric escape rates}
\begin{figure}
    \centering
    \includegraphics[width=0.5\textwidth]{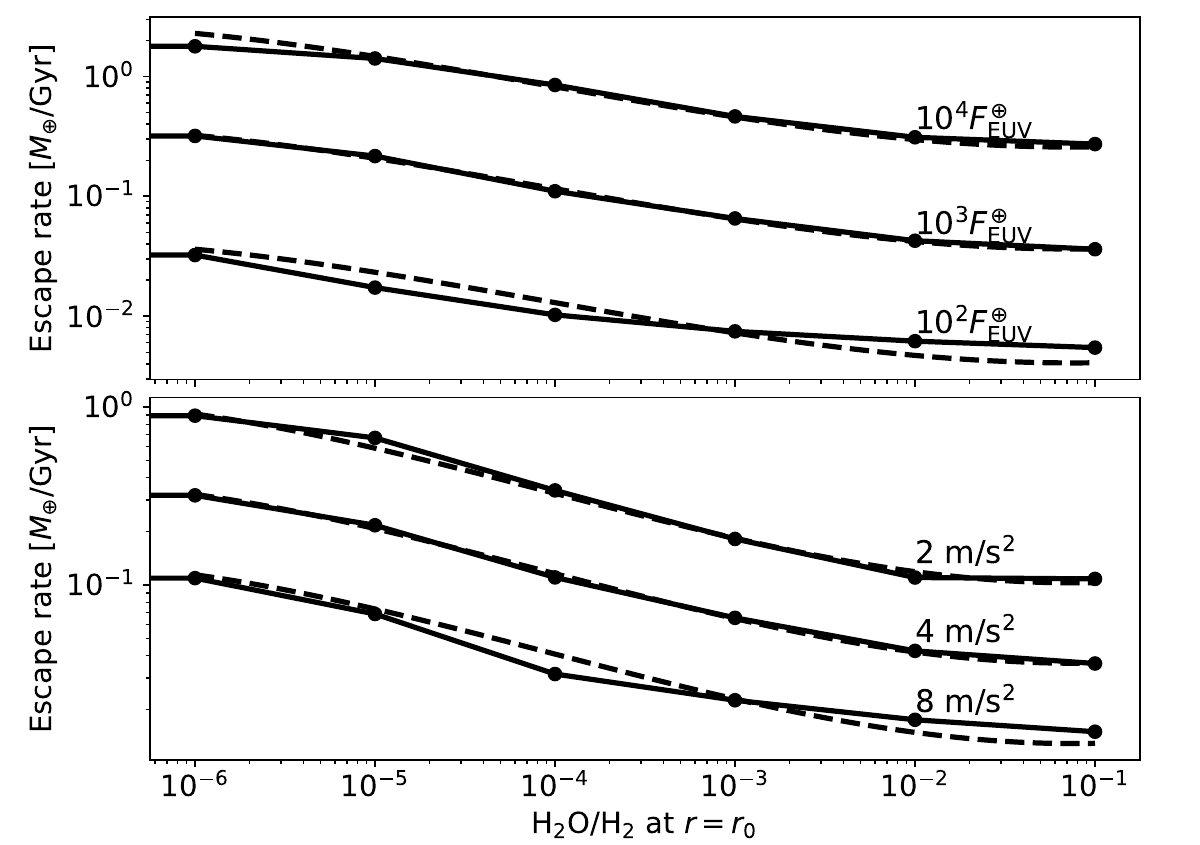}
    \vspace{-20pt}
    \caption{Total escape rate vs. \water/\htwo\ for different values of EUV irradiance at fixed surface gravity $g = 4$\,m\,s$^{-2}$ (top panel) and for different values of $g$ at a fixed EUV irradiance of $10^3 F_{\oplus}$ (bottom panel). $M_p$  is fixed at $5M_{\oplus}$.  The dashed lines are from Eqns. \ref{eqn:funct1}--\ref{eqn:funct3}.}
    \label{fig:escape-rates1}
\end{figure}

The escape rate $\dot{M}$ decreases with increasing basal H$_2$O/H$_2$ ratio due to enhanced radiative cooling and lower temperatures, regardless of gravity or EUV irradiance, with a 1 dex decrease at a basal H$_2$O/H$_2$ ratio of 0.1 relative to a pure H$_2$ atmosphere (Fig.~\ref{fig:escape-rates1}). This reduction in escape rate asymptotes at high H$_2$O/H$_2$ as the outflow becomes optically thick to IR radiation and H$_2$O photolysis happens deeper in the outflow due to slower advection.  At a given basal H$_2$O/H$_2$ ratio, the escape rate is approximately proportional to $F_{\rm EUV}$, and decreases with increasing $g$ approximately as $g^{-3/2}$, following the expected scaling for energy-limited escape \citep[e.g.,][]{Owen2019}. Detailed atmospheric profiles are shown in Appendix A. The escape rate as a function of H$_2$O/H$_2$ ratio, $F_{\rm EUV}$, $g$, and $M_p$ can be approximated by the following relation:
\begin{equation}
    \dot{M}\simeq \dot{M}_{\rm ref}\left(\frac{F_{\rm EUV}}{10^{3}F_{\rm EUV}^{\oplus}}\right)^{a_\mathrm{EUV}}\left(\frac{g}{4\,\mathrm{m/s^{2}}}\right)^{-3/2}\left(\frac{M_p}{5M_{\oplus}}\right)^{1/2},
\label{eqn:funct1}
\end{equation}
\vspace{-7pt}
\begin{equation}
\begin{split}
    \mathrm{log}_{10}\dot{M}_{\rm ref}\,[\mathrm{M_{\oplus}/Gyr}]=0.01042(\mathrm{log}_{10}r)^{3}+0.1257(\mathrm{log}_{10}r)^{2} \\
    +0.2429\mathrm{log}_{10}r-1.311,
\end{split}
\label{eqn:funct2}
\end{equation}
\vspace{-7pt}
\begin{equation}
    a_{\rm EUV}=-0.05\mathrm{log}_{10}\left(\frac{F_{\rm EUV}}{10^{3}F_{\rm EUV}^{\oplus}}\right)+0.9,
\label{eqn:funct3}
\end{equation}
where $F_{\rm EUV}^{\oplus}$ is the EUV flux at Earth and $r$ is the basal H$_2$O/H$_2$ ratio. $a_{\rm EUV}$ decreases with increasing $F_{\rm EUV}$ because as temperature increases, radiative cooling becomes more effective \citep{Yoshida+2021}. This function may become invalid under thermally instable conditions \citep{Kubyshkina+2018}, extremely high EUV flux and deep gravitational potential conditions such as those of hot Jupiters where atomic cooling is effective \citep{Murray-Clay2009,Salz+2016}, and in the diffusion-limited regime \citep{Hunten1973}.

\subsection{\water/\htwo-dependent evolution of sub-Neptunes}
\begin{figure}
    \centering
    \includegraphics[width=0.5\textwidth]{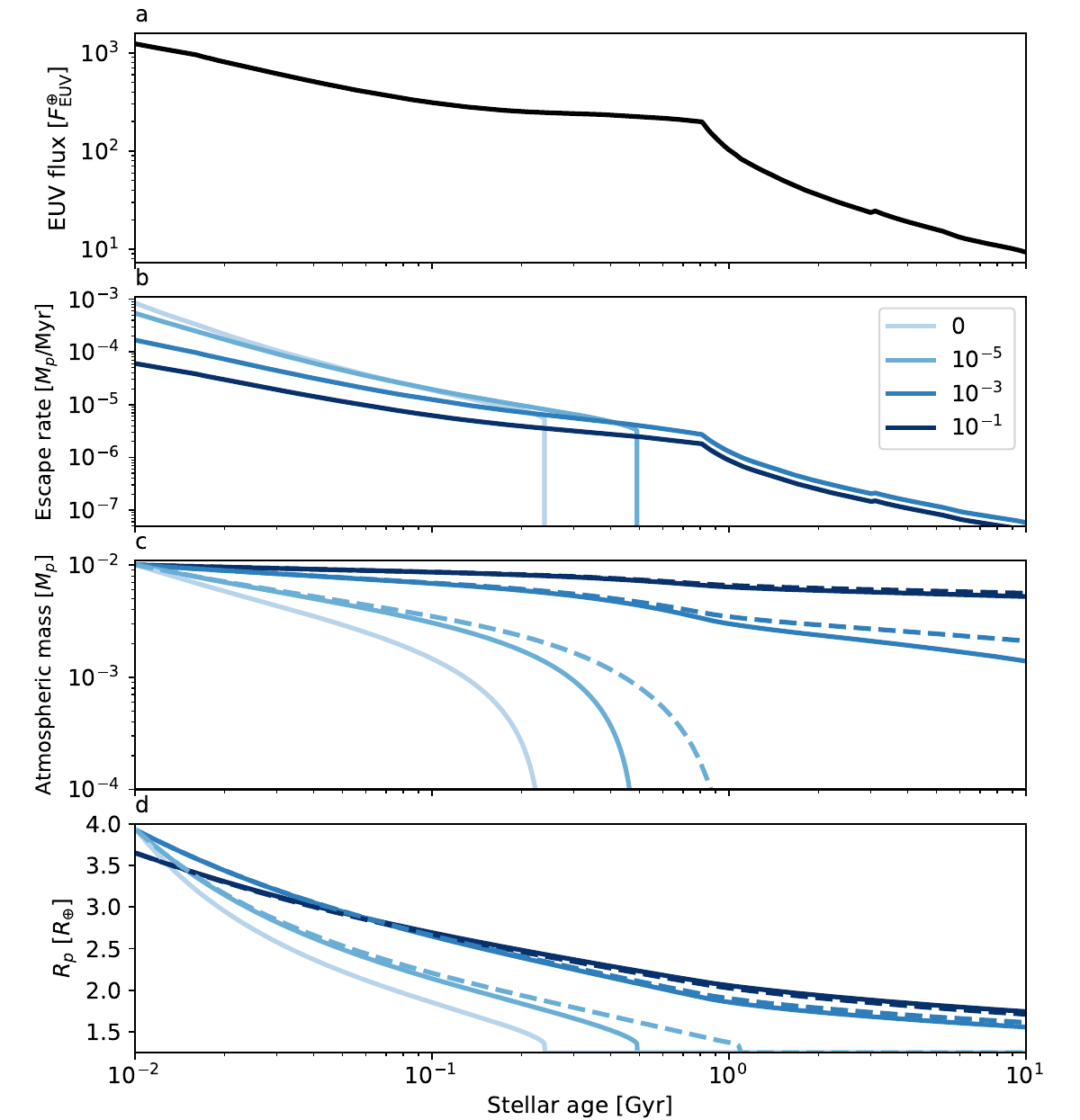}
    \vspace{-20pt}
    \caption{Evolution of EUV flux (top panel, a), escape rate (b), envelope mass (c), and planet radius (d) for a planet with $M_p = 2.5M_{\oplus}$, an initial envelope mass fraction of 1\%, and different initial values of \water/\htwo\ on an orbit with a main-sequence stellar irradiance equal to the terrestrial value. Solid (dashed) lines represent results without (with) mass fractionation between H and O.  The mass fractionation results have been omitted from panel b for clarity.}
    \label{fig:scenario}
\end{figure}

Figure~\ref{fig:scenario} shows the evolutionary tracks of atmospheres with different H$_2$O/H$_2$ ratios for $M_p = 2.5$\mearth\ and a 1\% initial envelope mass fraction, on an orbit where the main sequence stellar flux is equivalent to the solar constant. High H$_2$O/H$_2$ ratios extend atmospheric lifetimes by reducing escape rates through radiative cooling (Fig. \ref{fig:scenario}b). For H$_2$O/H$_2$ ratios above 10$^{-3}$, envelopes persist for at least 10 Gyr, whereas atmospheres with lower H$_2$O/H$_2$ ratios are lost within the period (Fig. \ref{fig:scenario}c). If mass fractionation between H and O is included (Appendix \ref{sec:fractionation}), atmospheric lifetimes slightly increase as the H$_2$O/H$_2$ ratio increases (dashed lines in Fig.~\ref{fig:scenario}c). Planets that retain significant envelopes have radii $\gtrsim 2R_{\oplus}$ (Fig. \ref{fig:scenario}d). 

\begin{figure}
    \centering
    \includegraphics[width=\linewidth]{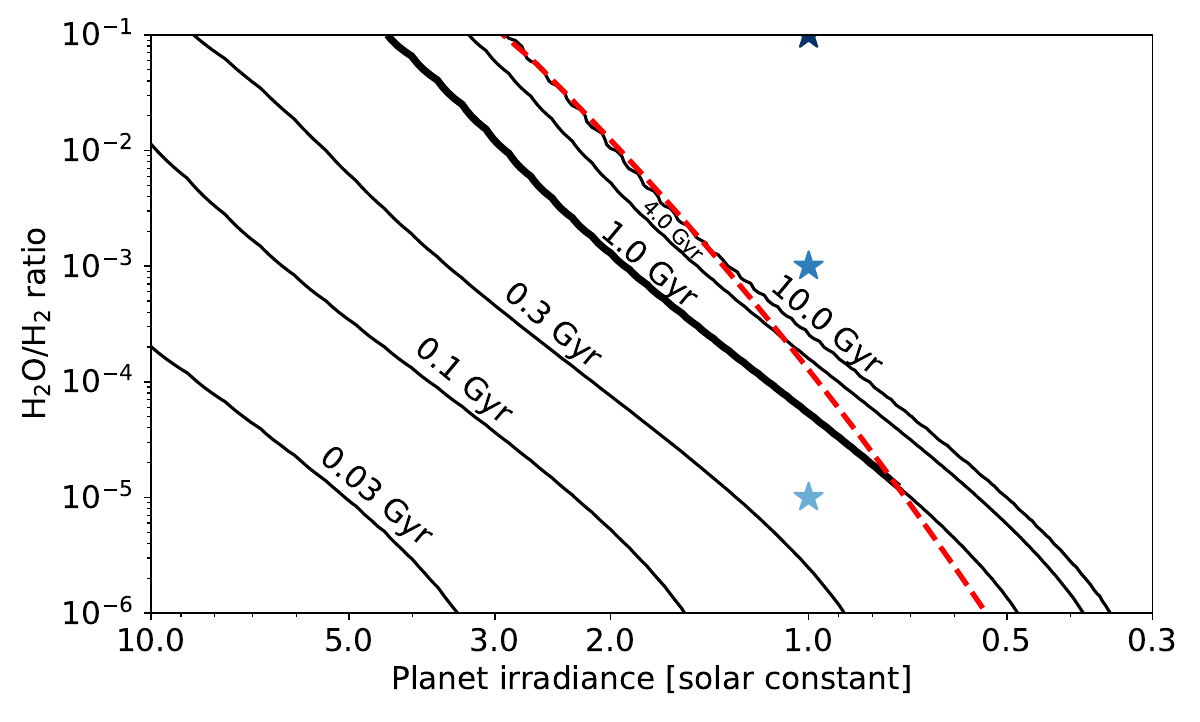}
    \vspace{-25pt}
    \caption{Lifetime of envelopes with initial mass fraction of 1\% on $M_p = 2.5M_{\oplus}$ sub-Neptunes as a function of stellar irradiance and H$_2$O/H$_2$. The dashed line is the saturation-limited H$_2$O/H$_2$ ratio in the upper atmosphere (see text and Appendix C). The stars represent the scenarios in Fig.~\ref{fig:scenario}.}
    \label{fig:irrad-water}
\end{figure}

Figure~\ref{fig:irrad-water} shows envelope lifetime as a function of stellar irradiance and the H$_2$O/H$_2$ ratio, where the lifetime is defined time when the atmospheric mass falls below $10^{-4}M_p$.  The three scenarios in Fig. \ref{fig:scenario} with \water/\htwo\ $>0$ are plotted as stars. The timescale increases with \water/\htwo\ and decreases with irradiance. The lifetime is weakly dependent on the initial envelope mass due to the dependence of gravity on planet radius and hence on envelope mass (Fig.~\ref{fig:Timescale_Mint}).

\section{Discussion}
\label{sec:discussion}
Around low luminosity K- and M dwarfs, temperate planets in which \water\ condensation will limit the abundaence in their upper atmospheres  at $\lesssim0.3$ au, within the range probed by exoplanet surveys such as \kepler.  \citet{Gaidos2024c} showed that the radius distribution of close-in planets around single M dwarfs, like that of solar-type stars, includes two peaks representing super-Earths and sub-Neptunes (see also \citealt{Hirano2018,Cloutier2020,VanEylen2021}).  Using stellar-rotation based ages, \citet{Gaidos2023b} found that the younger half of the sample (mean age of $\sim$2.5 Gyr) contains more sub-Neptunes relative to super-Earths than the older half (mean age of $\sim$6 Gyr).  This trend is similar to that found around solar-type stars \citep[e.g.,][]{Berger2020}, and expected from models in which sub-Neptunes lose their envelopes to EUV-driven escape and evolve into super-Earths.  Unexpectedly, \citet{Gaidos2024c} found that the disappearance of sub-Neptunes occurs primarily at larger distances and lower irradiance, near the point where \water\ is expected to condense in the atmospheres of temperate planets.  They speculated that the disappearance of cooler sub-Neptunes over Gyr timescales could be due to the collapse of steam-dominated atmospheres, or the accelerated loss of H-rich atmospheres from which \water\ has condensed.  

As shown in Fig.~\ref{fig:irrad-water}, envelopes are rapidly ($\ll$1 Gyr) lost from planets that are highly irradiated ($S>5S_{\oplus}$) and/or are \water-poor.  The dashed curve is the upper atmosphere saturation-limited \water\ concentration as a function of irradiance (i.e., planet equilibrium temperature) predicted by a simple model  of a $H_2$-dominated radiative-convective atmosphere with a Bond albedo equal to that of Neptune (see Appendix \ref{sec:water} for details).  The location of the curve with respect to the  contours of envelope lifetime suggests that a sub-Neptune irradiated at greater than the solar constant would retain its envelope and large radius for $\gtrsim$4 Gyr due to cooling by H$_2$O.  The same object at greater distances and lower irradiance would lose its envelope in $<4$ Gyr, possibly explaining the paradoxical findings of \citet{Gaidos2024c}.  \footnote{At \water/\htwo\ exceeding $0.1$ the metallicity is $>100$, the molecular weight is more than doubled, and due to that and increased cooling planets are less likely to have extended sub-Neptune-like atmospheres \citep{Linssen2024}.}  Cooling by other, less condensible molecules (e.g., CO, CH$_4$) will be important at low \water/\htwo\ (see below), such that the contours in Fig. \ref{fig:irrad-water} should become vertical, i.e., envelopes can persist at yet further distances. 

\begin{figure}
    \centering
    \includegraphics[width=0.5\textwidth]{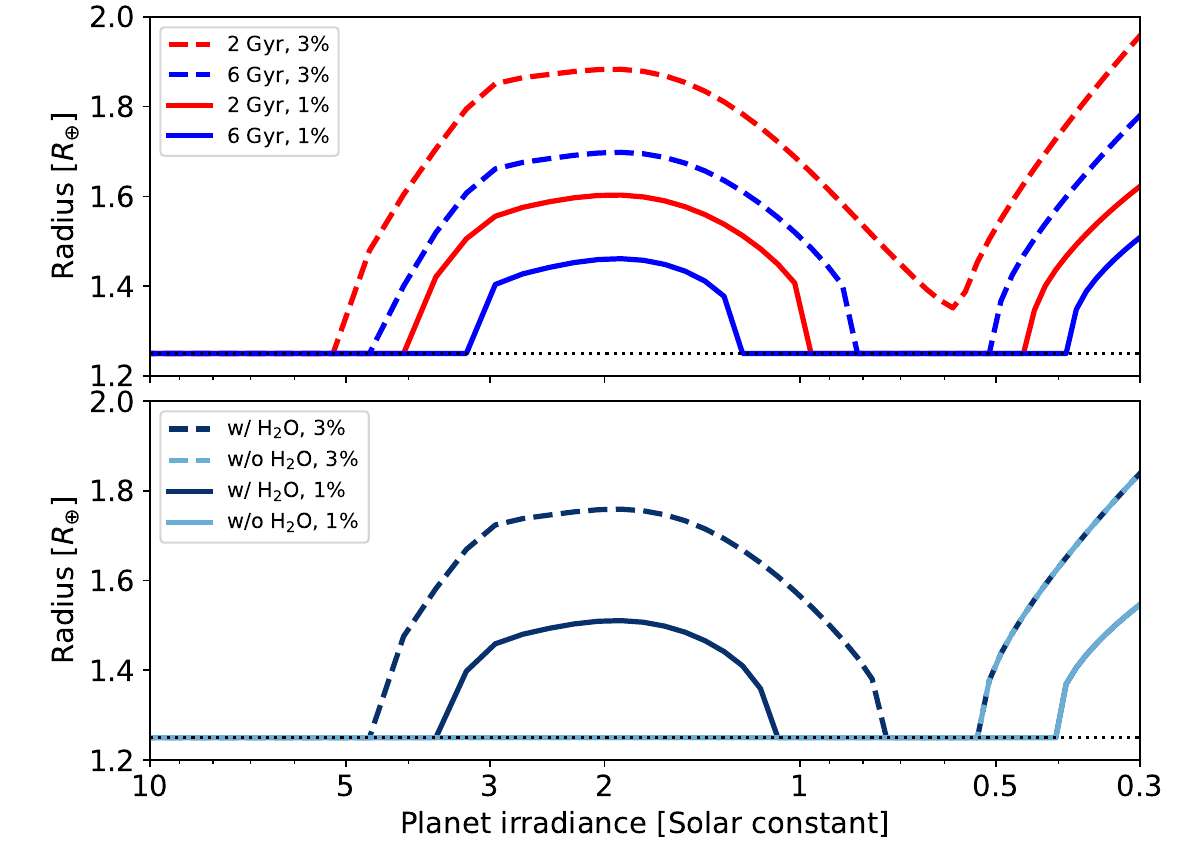}
    \vspace{-20pt}
    \caption{Top (a): Planet radius vs. irradiance for sub-Neptunes with a mass of $2.5M_{\oplus}$ and initial envelope mass fractions of 1\% (solid lines) and 3\% (dashed lines), at 2 and 6 Gyr.  H$_2$O/H$_2$ is set to the saturation limit (dashed line in Fig. \ref{fig:irrad-water}) up to a maximum of 0.1.  Bottom (b): Same for sub-Neptunes at 4 Gyr with \water\ as prescribed above, vs. those having envelopes without \water.}
    \label{fig:waterline}
\end{figure}

Fig.~\ref{fig:waterline}a demonstrates that, even if envelopes survive, the saturation-limited abundance of \water\ affects the radii and hence detectability by transit of sub-Neptunes observed at ages typical of \kepler\ target stars \citep[e.g.,][]{Dressing2015,Ment2023}.  It shows the radius of 2.5-\mearth\ sub-Neptunes vs. irradiance, for envelope mass fractions of 1\% and 3\% and \water/\htwo\ set by the saturation limit, after 2 Gyr and 6 Gyr of evolution.  At a given age, radii increase with decreasing (EUV) irradiance (due to increasing distance) until reaching an age- and envelope mass fraction-dependent maximum at twice the solar constant.  Thereafter the radius declines as decreasing (bolometric) irradiance and equilibrium temperatures suppress \water\ and its cooling of the upper atmosphere.  Below an irradiance of $\sim 50$\% the solar constant, cooling by \water\ is negligible and radii again increase with declining EUV flux.   

In water-poor systems, inefficient cooling will render primordial H/He envelopes around rocky cores more susceptible to hydrodynamic escape (Fig. \ref{fig:waterline}b) as well as erosion by stellar winds.  Infrared spectroscopy by \spitzer\ and \jwst\ has revealed that the chemistry of many inner protoplanetary disks around very low mass stars (VLMS, $\lesssim0.2$\,\msun) is consistent with a high C/O ratio \citep{Najita2013,Tabone2023,Arabhavi2024,Kanwar2024,Long2025}.  Sub-Neptunes arising from such disks could have high C/O, \water-poor envelopes.  Figure \ref{fig:waterline}b shows that, on orbits with super-terrestrial irradiance ($<$0.06\,au for an M2-type host) where planets are readily detected by transit surveys, such envelopes will completely escape by 4 Gyr.  While temperate, C/O-rich sub-Neptunes are predicted to contain abundant CH$_4$ \citep{Moses2013}, this molecule is a less efficient radiator and susceptible to UV photodissociation \citep{Yoshida2024,Yoshida2024b}.  CO from disequilibrium mixing \citep{Fortney2020} could be a source of O for the formation of \htwo\ and OH, but its high molecular weight will limit its altitude distribution to the turbopause, where CO-dissociating far-UV photons are already absorbed by overlying CH$_4$.  Thus planets forming from high C/O disks will experience accelerated  evaporation of their envelopes, which might explain the paucity of sub-Neptunes relative to rocky planets around VLMS compared to their FGK counterparts \citep{Ment2023}.    Not all disks around VLMS are O-poor \citep{Xie2023} and such exceptions could give rise to a surviving population of \water-rich (or at least metal-rich) sub-Neptunes, e.g. GJ\,1214b \citep{Ohno2025}.  

In the same vein, protoplanetary disks around stars in binary systems with separations much less than $\sim$100\,au are tidally truncated to a few au and are shorter lived \citep[see ][and references therein]{Zagaria2023}.  In these cases, the removal of icy, O-rich (\water\ and CO$_2$) solids in the outer disk that would otherwise drift inward could lead to O- and \water-poor planets in the inner disk.  These \water-poor objects would experience more rapid envelope escape, possibly explaining the deficit of detectable planets in such systems \citep{Kraus2016}, and in particular lack of sub-Neptunes relative to super-Earths \citep{Sullivan2023}.

The predicted decline in the abundance of \water\ at irradiances below $\sim$3 times the solar constant (Fig. \ref{fig:irrad-water}) can be detected by \jwst\ using the method of transmission spectroscopy during transits \citep[e.g.][]{Benneke2024,Holmberg2024}; such observations probe atmospheric composition at the 1--10 mbar level \citep{Beichman2018} .  All else being equal, indicators of atmospheric escape, e.g., anomalous absorption in H Lyman-$\alpha$ or the He \, triplet at 1083\,nm should be anti-correlated with the presence of \water.  

One caveat of our work is that we show outcomes for specific envelope mass fractions, whereas a population of sub-Neptunes would presumably have a distribution of values.  Nevertheless, the overall trends with \water/\htwo, irradiation, and stellar mass should hold.  Another is that the saturation concentration of \water\ at the radiative-convective boundary should be considered an upper limit to the value in the upper atmosphere.  The lower simulation boundary where \water/\htwo is fixed is at a number density of $\sim10^{13}$\,cm$^{-3}$ and an altitude above the turbopause where the atmosphere ceases to be well mixed.  Based on the eddy diffusivity relation for solar-metallicity atmospheres in \citet{Charnay2015} and a molecular diffusivity for \htwo\ \citep{Suarez-Iglesias2015}, the turbopause is predicted to lie at a pressure altitude of $8.7 \times 10^{-5}$\,bar or a density of $2.3 \times 10^{15}$\,cm$^{-3}$.  This corresponds to several scale heights below the base of the thermosphere, and could mean that \water\ is significantly depleted by diffusion.    \water\ is also photo-dissociated by FUV radiation, but in a H-rich atmosphere the back reaction is efficient, and there should be an equilibrium between \water\ and the principle photodissociation product O$_2$ \citep[e.g.,][]{Kawamura+2024}.  An O$_2$-rich layer will form on top of the \water-rich atmosphere below and protect it from FUV radiation (inset of top panel of Fig. \ref{fig:cooling}).  Finally, above a critical \water/\htwo ($10^{-2}$ for solar metallicity and higher for super-solar) , moist convection will be inhibited \citep{Guillot1995,LeConte2017}, altering the pressure-temperature profile. 

\section{Conclusions}
\label{sec:conclusions}
The hydrodynamic escape of H and He thought to be responsible for the evolution of gas-rich sub-Neptunes into rocky-dominated super-Earths is governed by the stellar EUV energy absorbed in the upper atmosphere relative to the energy of escape from planetary gravity.  This balance is modulated by IR and line emission by molecules and ions within the flow.  We use a hydrodynamic escape model of a spherically symmetric, EUV-driven wind to demonstrate that the abundance of \water\ in the upper atmosphere has a significant effect on escape rates.  Specifically, we find that:
\begin{itemize}
\item An increase in the \water\ abundance from $10^{-6}$ to $10^{-1}$ leads to a one dex decrease in escape rate, irrespective of surface gravity and XUV flux.
\item Sub-Neptunes just outside an irradiance level, approximately the solar constant, where saturation severely limits the \water\ concentration in the upper atmosphere will experience rapid decline in their radii relative to equivalent planets slightly closer to their host stars, potentially explaining the evolutionary trends inferred by \citet{Gaidos2024c}.
\item Sub-Neptunes that form around very low mass stars ($\lesssim$0.2\msun), which tend to have protoplanetary disks with \water-poor chemistries consistent with a high C/O ratio, are likely to experience more rapid escape, a possible explanation for their paucity around such stars.
\item A spectroscopic investigation of temperate sub-Neptune envelopes, i.e. by \jwst, should reveal a marked decline in water content with lower irradiance, correlating with the paucity of such objects around older stars.
\end{itemize}

\begin{acknowledgements}
TY was supported by JSPS KAKENHI Grant Nos. 23KJ0093 and 23H04645.  EG was supported by NASA Awards 80NSSC20K0957 (Exoplanets Research Program) and 80NSSC21K0597 (Interdisciplinary Consortia for Astrobiology Research) and as a Gauss Professor at the University of G\"{o}ttingen by the Nieders\"{a}chsische Akaemie der Wissenschaften. 
\end{acknowledgements}

\bibliographystyle{aa} 
\bibliography{references_master} 

\begin{appendix}
\section{Atmospheric Profiles}
\label{sec:profiles}
\begin{figure}
    \centering
    \includegraphics[width=0.5\textwidth]{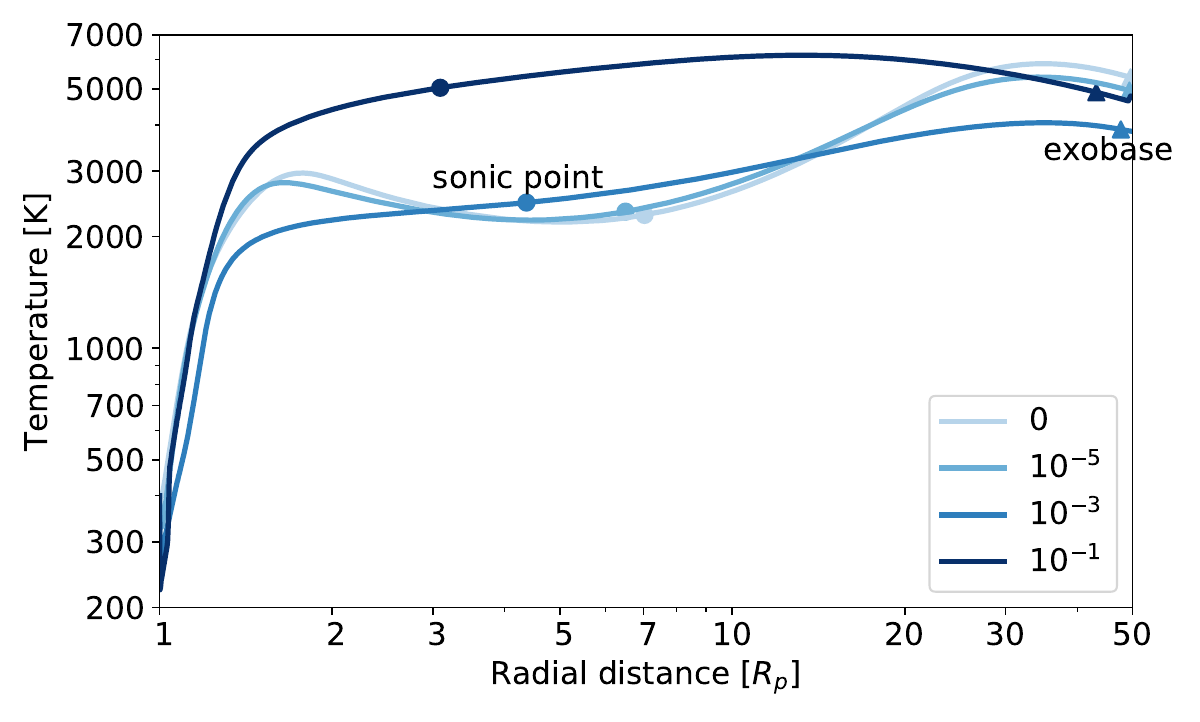}
    \vspace{-25pt}
    \caption{Temperature profile (radial coordinate in units of planet radius) in a hydrodynamic flow for different lower boundary values \water/\htwo\ with a stellar EUV flux of $1000F_{\rm EUV}^{\oplus}$, $M_{p}=5M_{\oplus}$, and $g =4\,\mathrm{m\,s^{-2}}$, where $F_{\rm EUV}^{\oplus}$ is the present EUV flux at Earth.  Sonic points are marked by filled circles and the exobase is marked by triangles.}
    \label{fig:temp}
\end{figure}

\begin{figure}
    \centering
    \includegraphics[width=0.5\textwidth]{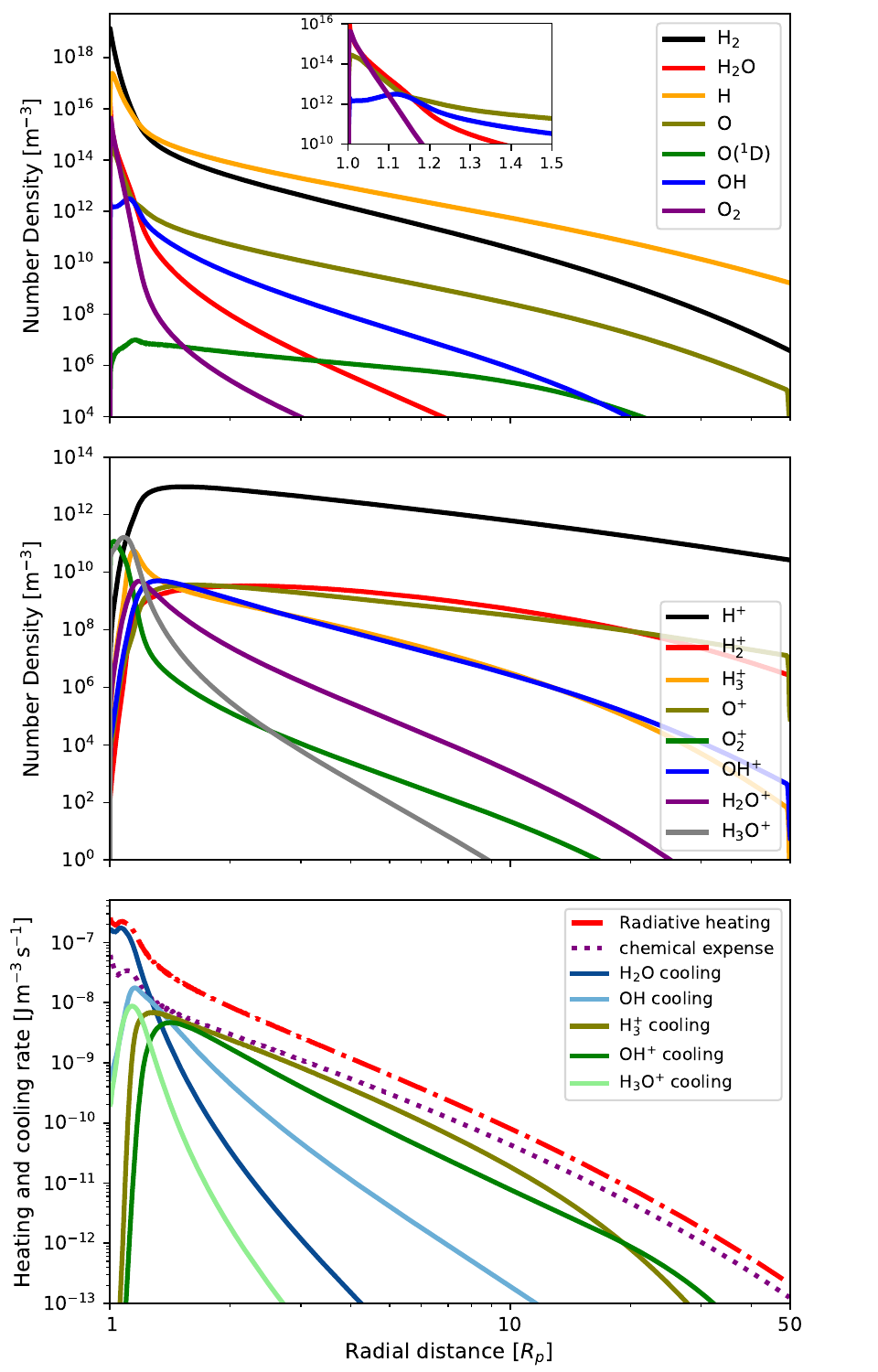}
    \caption{Top: Number density profiles of neutral species with a basal H$_2$O/H$_2$ of 0.001, stellar EUV flux of $1000F_{\rm EUV}^{\oplus}$, $M_{p}=5M_{\oplus}$, and $g =4\,\mathrm{m\,s^{-2}}$.  The inset panel shows details near the base of the flow.  Middle: Predicted number density profiles of ions.  Bottom: Contribution of these species to cooling rates. ``Radiative heating" is the heating rate by adsorption of X-rays and UV, and ``chemical expense" is the energy used by chemical reactions such as photolysis.}
    \label{fig:cooling}
\end{figure}

Fig.~\ref{fig:temp} shows the dependence of the outflow temperature profile on basal H$_2$O/H$_2$ subjected to stellar EUV $10^3\times$ the present terrestrial value, with $M_{p}=5M_{\oplus}$ and $g=4\,\mathrm{m\,s^{-2}}$. Fig. \ref{fig:cooling} shows profiles of the number density, heating, and cooling rate in a flow with the same parameters, except only for a basal H$_2$O/H$_2$ ratio of $10^{-3}$,  In the lower region of the flow ($r \lesssim 2R_{p}$), the temperature decreases with increasing basal H$_2$O/H$_2$ ratio up to $\sim 0.01$ as a result of greater radiative cooling.   As shown in Fig.~\ref{fig:cooling}, H$_2$O is the primary coolant in this lower region.  Along the outflow, H$_2$O is photolyzed to H, O, OH, and various ions. As a result, cooling by photochemical products, especially H$_{3}^{+}$ and OH$^{+}$, is significant at higher altitudes.   As the basal H$_2$O/H$_2$ ratio approaches $\sim$0.1, the mean molecular weight of the flow increases, reducing the speed of the outflow and the contribution of adiabatic cooling, and temperatures in the lower region of the flow begin to increase. This dependence on the mean molecular weight is discussed in detail by \citet{Yoshida+2020}.

\section{Mass fractionation between H and O}
\label{sec:fractionation}
\begin{figure}
    \centering
    \includegraphics[width=0.5\textwidth]{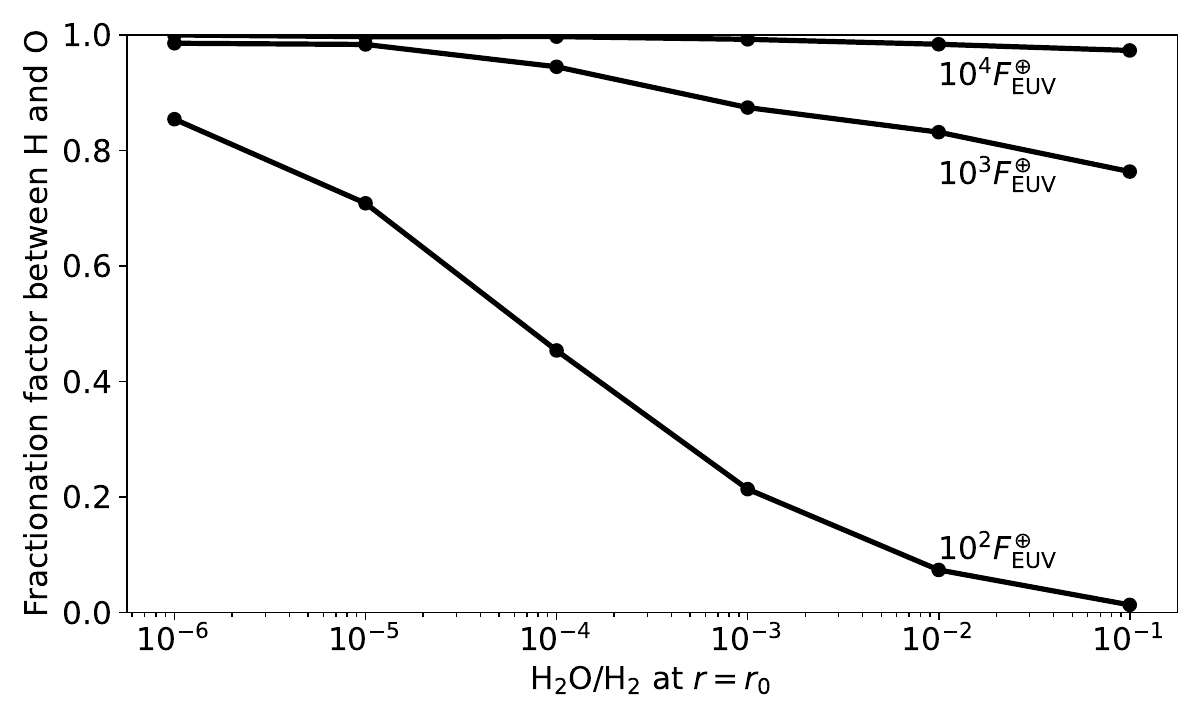}
    \vspace{-25pt}
    \caption{Fractionation factor between H and O for different values of interior \water/\htwo\ and stellar EUV flux with a surface gravity of $4\,\mathrm{m\, s^{-2}}$ and planet mass of $5M_{\oplus}$.}
    \label{fig:frac}
\end{figure}

Fig. \ref{fig:frac} shows the fractionation factor between H and O, defined as $(F_{\rm O}/F_{\rm H})/(n_{\rm O}(r_0)/n_{\rm H}(r_0))$, where $F_{i}$ and $n_{i}(r_0)$ are the escape rate and the total number density at the lower boundary $r_0$ of element $i$, respectively. This ratio decreases (fractionation increases) as the H$_2$O/H$_2$ ratio increases or as the EUV flux decreases, due to the reduced H escape flux and less efficient O drag.

\section{Escape timescale as a function of initial atmospheric mass and stellar flux}
\label{sec:Escape-timescale}

\begin{figure}
    \centering
    \includegraphics[width=0.5\textwidth]{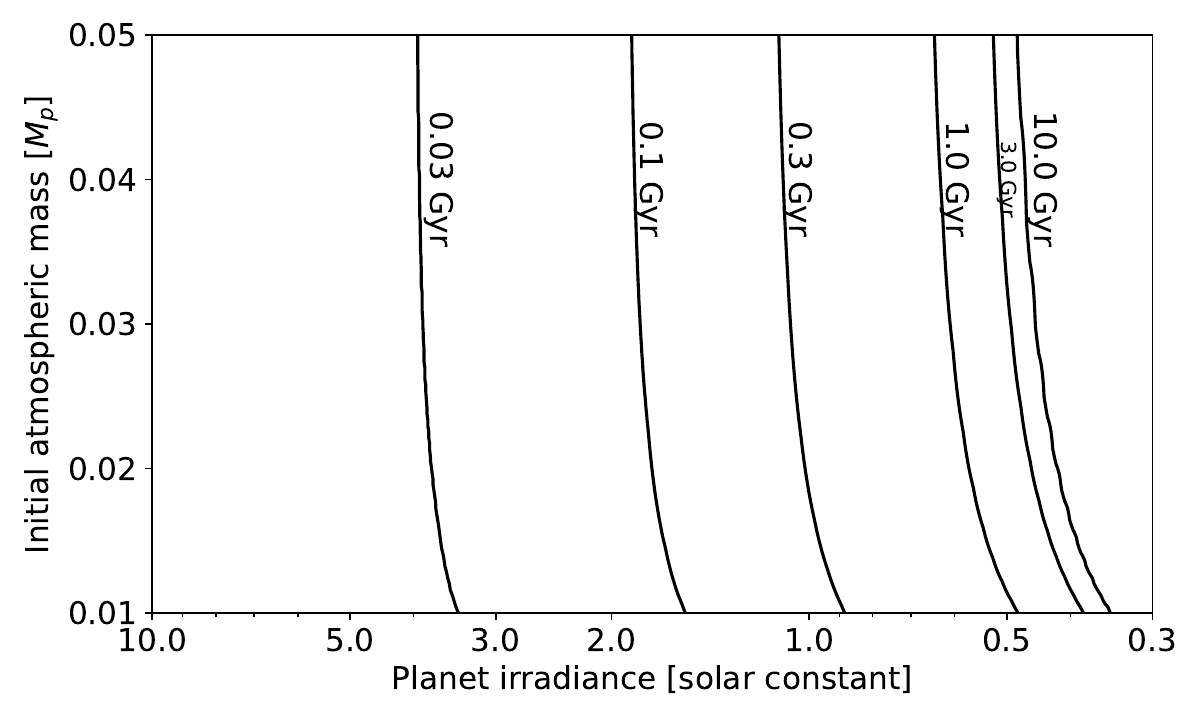}
    \vspace{-25pt}
    \caption{Escape timescale of pure H$_2$ atmospheres as a function of stellar irradiance and initial envelope mass with a planet mass of $2.5M_{\oplus}$. }
    \label{fig:Timescale_Mint}
\end{figure}

Fig.~\ref{fig:Timescale_Mint} shows the escape timescale of pure H$_2$ atmosphere as a function of stellar flux and initial atmospheric mass with a planetary mass of $2.5M_p$, where the timescale is defined as the duration required for the atmospheric mass to reach $10^{-4}M_p$. The timescale exhibits weak dependence on the initial atmospheric mass because the escape rate increases with initial atmospheric mass due to planetary inflation and the resulting decrease in gravitational acceleration at the base of the outflow.

\section{Model details}
\label{sec:details}
\subsection{Basic equations}

We solve the fluid equations of continuity, momentum and energy for a multi-component gas, assuming a spherically symmetric flow:
 	\begin{equation}
 		\frac{\partial n_{i}}{\partial t}+\frac{1}{r^{2}}\frac{\partial (n_{i} u_{i}r^{2})}{\partial r}=\omega_{i},
 	\end{equation}
 	\begin{equation}
 		\frac{\partial u_{i}}{\partial t}+u_{i}\frac{\partial u_{i}}{\partial r}=-\frac{1}{\rho_{i}}\frac{\partial p_{i}}{\partial r}-\frac{GM_{p}}{r^{2}}+\sum_{\substack{j}}(u_{j}-u_{i})\frac{n_{j}}{m_{i}}\mu_{ij}k_{ij},
 	\end{equation}
 	\begin{equation}
 		\frac{\partial}{\partial t}\left[\rho \left(\frac{1}{2}u^{2}+E\right)\right]+\frac{1}{r^{2}}\frac{\partial}{\partial r}\left\{\left[\left(\frac{1}{2}u^{2}+E+\frac{p}{\rho}\right)\rho u \right]r^{2}\right\} =-\frac{GM_{p}}{r^{2}}\rho u+q,
 	\end{equation}
    where $t$ is the time, $r$ is the distance from the planet's center, $n_{i}, \rho_{i}, u_{i}, p_{i}$, and $\omega_{i}$ are the number density, mass density, velocity, partial pressure, and production rate of species $i$, respectively, $G$ is the gravitational constant, $M_p$ is the mass of the planet, $\rho$, $p$, and $E$ are the total mass density,  total pressure, and total specific internal energy, respectively, $u$ is the mean gas velocity, $\mu_{ij}$ is the reduced molecular mass between species $i$ and $j$, $k_{ij}$ is the momentum transfer collision frequency that follows
	\begin{equation}
 		k_{ij} = \frac{k_{\rm B}T}{\mu_{ij}b_{ij}},
 	\end{equation}
	where $k_{\mathrm{B}}$ is the Boltzmann constant and $T$ is the temperature.  $b_{ij}$ is the binary diffusion coefficient given by
	\begin{equation}
  		b_{ij} = \begin{cases}
   			1.96\times 10^{6}\frac{T^{1/2}}{\mu_{ij}^{1/2}}\hspace{32.5pt}\mathrm{for\,neutral \mathchar`- neutral\,pair} \\
			4.13\times 10^{-8}\frac{T}{\mu_{ij}^{1/2}\alpha^{1/2}}\hspace{14pt}\mathrm{for\,neutral \mathchar`- ion\,pair} \\
			8.37\times 10^{-5}\frac{T^{5/2}}{\mu_{ij}^{1/2}}\hspace{27pt}\mathrm{for\,ion \mathchar`- ion\,pair}
  		\end{cases}
	\end{equation}
	where $\alpha$ is the polarizability (\citealt{Banks+1973}; \citealt{Munoz2007}). Here, $b_{ij}$, $T$, $\mu_{ij}$ and $\alpha$ are expressed in $\mathrm{cm}^{-1}\mathrm{s}^{-1}$, $\mathrm{K}$, $\mathrm{g}$ and $\mathrm{cm^{3}}$, respectively. The total internal energy is given by	
	\begin{equation}
		\rho E= \frac{1}{n}\left(\sum_{\substack{i}}\frac{n_{i}}{\gamma_{i}-1}\right)p,
 	\end{equation}
	where $n$ is the total number density and $\gamma_{i}$ is the ratio of specific heat of species $i$. The net heating rate is given by 
	\begin{equation}
		q=q_{\mathrm{abs}}-q_{\mathrm{ch}}-q_{\mathrm{rad}},
 	\end{equation}
	where $q_{\mathrm{abs}}$ is the heating rate by X-ray and UV absorption (see \ref{sec:radiation}), $q_{\mathrm{ch}}$ is the rate of net chemical expense of energy (see \ref{sec:chemistry}), and $q_{\mathrm{rad}}$ is the radiative cooling rate by IR active molecules (see \ref{sec:radiation}).

\subsection{Chemistry}
\label{sec:chemistry}
	A total of 93 chemical reactions are considered for 15 atmospheric components: H$_{2}$, H$_{2}$O, H, O, O($^{1}$D), OH, O$_{2}$, H$^{+}$, H$_{2}^{+}$, H$_{3}^{+}$, O$^{+}$, O$_{2}^{+}$, OH$^{+}$, H$_{2}$O$^{+}$, and H$_{3}$O$^{+}$ (Table A1 and A2). 
	
	The photolysis rate $f_{\mathrm{ph},i}$ is given by
	\begin{equation}
		f_{\mathrm{ph},i}=\int_{\lambda}n_{i}\sigma_{i}(\lambda)I(\lambda)d\lambda,
 	\end{equation}
	where $\sigma_{i}(\lambda)$ is the photodissociation/photoionization cross section at wavelength $\lambda$ for species $i$ and $I(\lambda)$ is the incident X-ray/UV photon flux at wavelength $\lambda$. The energy consumed per unit time by this photolysis $q_{\mathrm{ph}, i}$ is given by
	\begin{equation}
		q_{\mathrm{ph}, i}=f_{\mathrm{ph},i}\frac{hc}{\lambda_{\mathrm{th},i}},
 	\end{equation}
	where $\lambda_{\mathrm{th},i}$ is the threshold wavelength for the photolysis of species $i$, $h$ is the Planck constant, and $c$ is the speed of light in vacuum. We adopt the photodissociation and photoionization cross sections provided by “PHoto Ionization/Dissociation RATES” (\citealt{Huebner+2015}; \url{http://phidrates.space.swri.edu}).
	
	In addition to photolysis reactions, this study considers bimolecular reactions. The bimolecular reaction rate $f_{\rm R}$ can be written as
	\begin{equation}
		f_{\mathrm{R}}=k_{\mathrm{R}}n_{i}n_{j},
 	\end{equation}
	where $n_{i}$ and $n_{j}$ represent the number densities of reactant $i$ and $j$, respectively, and $k_{\rm R}$ is the corresponding reaction rate coefficient. We use the reaction rate coefficients provided by ``The UMIST Database for Astrochemistry 2012" (\citealt{Mcelroy+2013}; \url{http://udfa.ajmarkwick.net}). We neglect the formation of molecules that have more than one carbon atom because of the low densities in the outflow. The energy consumed per unit time by a bimolecular reaction is given by
	\begin{equation}
		q_{\mathrm{R}}=-f_{\mathrm{R}}\Delta E_{\mathrm{R}},
 	\end{equation}
	where $\Delta E_{\mathrm{R}}$ is the heat of reaction, which is positive for endothermic reactions and negative for exothermic reactions. In evaluating the heats of reaction, we make use of the enthalpies of formation listed in \citet{Le+2000}.  The energy consumption rate $q_{\mathrm{ch}}$ is calculated by summing the consumed amounts of energy by individual chemical reactions including photolysis.

\subsection{Radiative processes}
\label{sec:radiation}
	In order to calculate the radial profiles of rates of heating and photolysis in the atmosphere, we model the radiative transfer of parallel stellar photon beams in a spherically symmetric atmosphere by applying the method formulated by \citet{Tian+2005}. We consider X-ray and UV absorption by H$_{2}$, H$_{2}$O, H, O, OH, and O$_{2}$. The stellar beam is assumed to be absorbed by Beer's law. The energy deposition in a given layer along each path is then multiplied by the area of the ring (Fig. 6 in \citealt{Tian+2005}) to obtain the total energy deposited into the shell segment. The volumetric heating rate $q_{\mathrm{abs}}$ is given by the total energy absorbed per unit time by each shell divided by the volume of the shell. We adopt the X-ray and UV spectrum from 0.1 to 280 nm estimated for GJ\,832 as a representative M2 dwarf \citep{Loyd+2016}. The EUV flux at 1 au is taken to be $1.1\times 10^{-3}\,\mathrm{W m^{-2}}$, which is about 0.25 times that from the present Sun. To represent both the higher EUV luminosity typical of young stars and a range of semi-major axes, we scale the stellar spectrum to consider EUV fluxes up to $10^4$ times the Earth’s level.
	
	We consider radiative cooling by thermal line emission of H$_{2}$O, OH, H$_{3}^{+}$, OH$^{+}$, and H$_3$O$^+$. The radiative cooling rate due to a transition from energy level $i$ to $j$ of radiatively active molecular species $s$ is given by
	\begin{equation}
		q_{\mathrm{rad},s,ij}=n_{s,i}A_{ij}h\nu_{ij}\beta_{ij},
 	\end{equation}
	where $n_{s,i}$ is the population density in level $i$, $A_{ij}$ is the spontaneous transition probability, $h\nu_{ij}$ is the energy difference between level $i$ and level $j$, and $\beta_{ij}$ is the photon escape probability. The total radiative cooling rate by species $s$ is calculated by summing the radiative cooling rates of all the transitions. The population density $n_{s,i}$ is calculated under the assumption of local thermodynamic equilibrium as
	\begin{equation}
		n_{s,i}=n_{s}\frac{g_{i}}{Z}\mathrm{exp}\left(-\frac{E_{i}}{k_{\mathrm{B}}T}\right)
 	\end{equation}
	where $n_{s}$ and $Z$ are the total number density and partition function of species $s$, and $g_{i}$ and $E_{i}$ are the statistical weight and energy of level $i$. Approximating that half of the photons are emitted upward while the other half is emitted downward and then absorbed by dense, lower atmospheric layers, the bulk escape probability is evaluated by
	\begin{equation}
		\beta_{ij}(r)=\frac{1}{2}\int_{0}^{\infty}\phi(\nu,r)\,\mathrm{exp}\left[-\int_{r_{\mathrm{top}}}^{r}\alpha_{\mathrm{L}_{ij}}\phi(\nu,r)dr\right] d\nu
	\end{equation}
	where $r_{\mathrm{top}}$ is the radius of the outer boundary (the top of the atmosphere), $\alpha_{\mathrm{L}_{ij}}$ is the absorption coefficient integrated over frequency, and $\phi(\nu,r)$ is the line profile, $\alpha_{\mathrm{L}_{ij}}$ is given by
	\begin{equation}
		\alpha_{\mathrm{L}_{ij}}=\frac{c^{2}}{8\pi \nu_{ij}^{2}}n_{s,i}A_{ij}\left(\frac{n_{s,j}g_{i}}{n_{s,i}g_{j}}-1\right).
	\end{equation}
	Assuming a Doppler-broadened line profile, $\phi(\nu,r)$ is given by
	\begin{equation}
 		\phi(\nu,r)=\frac{1}{\sqrt{\pi}\Delta \nu_{\rm D}}\mathrm{exp}\left[-\frac{1}{(\Delta \nu_{\rm D})^{2}}\left(\nu - \nu_{0}-\frac{u_{s}(r)}{c}\nu_{0}\right)^{2}\right],
	\end{equation}
	where $u_{s}(r)$ is the velocity profile of radiative sources and $\nu_{0}$ is the rest-frame central frequency of the line profile; and $\Delta \nu_{\rm D}$ is the Doppler width given by
	\begin{equation}
		\Delta \nu_{\rm D}=\frac{\nu_{0}}{c}\sqrt{\frac{3k_{\rm B}T}{m_{s}}},
	\end{equation}
	where $m_{s}$ is the molecular mass of species $s$. The net radiative cooling rate $q_{\mathrm{rad}}$ is obtained by the sum of contributions by H$_{2}$O, OH, H$_{3}^{+}$, and OH$^{+}$.   We use the line data provided by the {\tt HITRAN} database (\citealt{Rothman+13}; \url{http://hitran.org}) and {\tt ExoMol} database (\citealt{Tennyson+2016}; \url{http://exomol.com}). We consider 31,333 transitions of H$_{2}$O and 4043 transitions of OH to cover over 99\% of total energy emission within the 100--3000 K temperature range. The emission from H$_{3}^{+}$, OH$^+$, and H$_3$O$^+$ are assumed to be optically thin throughout the flow, with all lines from the {\tt ExoMol} database \citep{Tennyson+2016}.

\subsection{Calculation method}
	The governing equations are solved by numerical time integration until the physical quantities reach their steady-state profiles.  These equations can be split into advection phases and non-advection phases. We employ the CIP method to solve for the advection phases \citep{Yabe+1991} to achieve numerical stability and accuracy for hydrodynamic escape simulations \citep{Kuramoto+2013}. After solving the advection phases, we solve the non-advection phases with a finite-difference approach. The non-advection phases of the energy equation are solved explicitly. Those of the continuity equations are solved with the semi-implicit method. Those of the momentum equations are solved implicitly using the quantities for the next time step that are obtained by the integration of the energy equation and the continuity equations. 
	
	1000 spatial grids in radial coordinates are taken with the grid-to-grid intervals exponentially increasing with $r$. The time step is defined so as to satisfy the CFL condition. For each parameter setup, we continue the time integration until reaching the steady state using the convergence condition described in \citet{Tian+2005}.
	
	The lower and upper boundaries are set to $r=R_{p},\,50\,R_{p}$, respectively, where $R_{p}$ is the radius of the planet. In each simulation run, the atmospheric density and temperature at the lower boundary are fixed. We assume that only H$_{2}$ and H$_{2}$O exist at the lower boundary. The number density of H$_{2}$ at the lower boundary is set to $1\times 10^{19}\,\mathrm{m^{-3}}$. The number density of H$_{2}$O is given as a parameter in each simulation run. The temperature at the lower boundary is set to 400 K.  This corresponds to the skin temperature, the asymptotic temperature at high altitudes of the upper atmosphere that is optically thin in the thermal-IR and transparent to shortwave radiation \citep{Catling+2017}. The other physical quantities at the lower and upper boundaries are estimated by linear extrapolations from the calculated domain. As the initial condition, we use the steady-state profiles of the pure \htwo\ atmosphere and add H$_{2}$O whose number density profiles are given by the hydrostatic structure. The initial velocity of H$_{2}$O is set at $1\times 10^{-5}$ m s$^{-1}$. 

\begin{table*}
	\centering
	\caption{Chemical reactions included in the model}
	\label{tab:A1}
	\begin{tabular}{llcll} 
	\hline
	No. & Reaction &  &   & Reaction rate $\mathrm{cm^{3}\,s^{-1}}$\\
	\hline
	R1 & $\mathrm{H}_{2}+h\nu$ & $\to$ & $\mathrm{H}+\mathrm{H}$\\
	R2 &                                        & $\to$ & $\mathrm{H}_{2}^{+}+\mathrm{e}$\\
	R3 &                                        & $\to$ & $\mathrm{H}^{+}+\mathrm{H}+\mathrm{e}$\\
	R4 & $\mathrm{H_{2}O}+h\nu$ & $\to$ & $\mathrm{H}+\mathrm{OH}$\\
	R5 &                                        & $\to$ & $\mathrm{H}_{2}+\mathrm{O(^{1}D)}$\\
	R6 &                                        & $\to$ & $\mathrm{O}+\mathrm{H}+\mathrm{H}$\\		
	R7 &                                        & $\to$ & $\mathrm{OH^{+}}+\mathrm{H}+\mathrm{e}$\\
	R8 &                                        & $\to$ & $\mathrm{O^{+}}+\mathrm{H_{2}}+\mathrm{e}$\\
	R9 &                                        & $\to$ & $\mathrm{H^{+}}+\mathrm{OH}+\mathrm{e}$\\
	R10 &                                        & $\to$ & $\mathrm{H_{2}O^{+}}+\mathrm{e}$\\
	R11 & $\mathrm{H}+h\nu$ & $\to$ & $\mathrm{H^{+}}+\mathrm{e}$\\
	R12 & $\mathrm{O}+h\nu$ & $\to$ & $\mathrm{O^{+}}+\mathrm{e}$\\
	R13 & $\mathrm{OH}+h\nu$ & $\to$ & $\mathrm{O}+\mathrm{H}$\\
	R14 &                                        & $\to$ & $\mathrm{OH^{+}}+\mathrm{e}$\\
	R15 &                                        & $\to$ & $\mathrm{O(^{1}D)}+\mathrm{H}$\\
	R16 & $\mathrm{O_{2}}+h\nu$ & $\to$ & $\mathrm{O}+\mathrm{O}$\\
	R17 &                                        & $\to$ & $\mathrm{O}+\mathrm{O(^{1}D)}$\\
	R18 &                                        & $\to$ & $\mathrm{O^{+}}+\mathrm{O}+\mathrm{e}$\\
	R19 &                                        & $\to$ & $\mathrm{O_{2}^{+}}+\mathrm{e}$\\					
	R20 & $\mathrm{H_{2}}+\mathrm{H_{2}}$ & $\to$ & $\mathrm{H_{2}}+\mathrm{H}+\mathrm{H}$ & $1.0\times 10^{-8}\mathrm{exp}(-84000/T)$\\
	R21 & $\mathrm{H_{2}}+\mathrm{H_{2}O}$ & $\to$ & $\mathrm{OH}+\mathrm{H_{2}}+\mathrm{H}$ & $5.8\times 10^{-9}\mathrm{exp}(-52900/T)$\\
	R22 & $\mathrm{H_{2}}+\mathrm{O_{2}}$ & $\to$ & $\mathrm{O}+\mathrm{O}+\mathrm{H_{2}}$ & $6.0\times 10^{-9}\mathrm{exp}(-52300/T)$\\
	R23 & $\mathrm{H_{2}}+\mathrm{O_{2}}$ & $\to$ & $\mathrm{OH}+\mathrm{OH}$ & $3.16\times 10^{-10}\mathrm{exp}(-21890/T)$\\			
	R24 & $\mathrm{H_{2}}+\mathrm{OH}$ & $\to$ & $\mathrm{O}+\mathrm{H_{2}}+\mathrm{H}$ & $6.0\times 10^{-9}\mathrm{exp}(-50900/T)$\\
	R25 & $\mathrm{H_{2}}+\mathrm{H}$ & $\to$ & $\mathrm{H}+\mathrm{H}+\mathrm{H}$ & $4.67\times 10^{-7}(T/300)^{-1}\mathrm{exp}(-55000/T)$\\
	R26 & $\mathrm{H_{2}}+\mathrm{H_{2}^{+}}$ & $\to$ & $\mathrm{H_{3}^{+}}+\mathrm{H}$ & $2.08\times 10^{-9}$\\
	R27 & $\mathrm{H_{2}}+\mathrm{H_{2}O^{+}}$ & $\to$ & $\mathrm{H_{3}O^{+}}+\mathrm{H}$ & $6.40\times 10^{-10}$\\
	R28 & $\mathrm{H_{2}}+\mathrm{O^{+}}$ & $\to$ & $\mathrm{OH^{+}}+\mathrm{H}$ & $1.70\times 10^{-9}$\\
	R29 & $\mathrm{H_{2}}+\mathrm{OH^{+}}$ & $\to$ & $\mathrm{H_{2}O^{+}}+\mathrm{H}$ & $1.01\times 10^{-9}$\\
	R30 & $\mathrm{H_{2}}+\mathrm{O}$ & $\to$ & $\mathrm{OH}+\mathrm{H}$ & $3.14\times 10^{-13}(T/300)^{2.7}\mathrm{exp}(-3150/T)$\\
	R31 & $\mathrm{H_{2}}+\mathrm{OH}$ & $\to$ & $\mathrm{H_{2}O}+\mathrm{H}$ & $2.05\times 10^{-12}(T/300)^{1.52}\mathrm{exp}(-1736/T)$\\
	R32 & $\mathrm{H_{2}}+\mathrm{O(^{1}D)}$ & $\to$ & $\mathrm{H}+\mathrm{OH}$ & $1.0\times 10^{-9}$\\	
	R33 & $\mathrm{H_{2}O}+\mathrm{H}$ & $\to$ & $\mathrm{OH}+\mathrm{H}+\mathrm{H}$ & $5.80\times 10^{-9}\mathrm{exp}(-52900/T)$\\
	R34 & $\mathrm{H_{2}O}+\mathrm{H^{+}}$ & $\to$ & $\mathrm{H_{2}O^{+}}+\mathrm{H}$ & $6.90\times 10^{-9}(T/300)^{-0.5}$\\
	R35 & $\mathrm{H_{2}O}+\mathrm{H_{2}^{+}}$ & $\to$ & $\mathrm{H_{2}O^{+}}+\mathrm{H_{2}}$ & $3.90\times 10^{-9}(T/300)^{-0.5}$\\
	R36 & $\mathrm{H_{2}O}+\mathrm{O^{+}}$ & $\to$ & $\mathrm{H_{2}O^{+}}+\mathrm{O}$ & $3.20\times 10^{-9}(T/300)^{-0.5}$\\
	R37 & $\mathrm{H_{2}O}+\mathrm{OH^{+}}$ & $\to$ & $\mathrm{H_{2}O^{+}}+\mathrm{OH}$ & $1.59\times 10^{-9}(T/300)^{-0.5}$\\
	R38 & $\mathrm{H_{2}O}+\mathrm{H_{2}^{+}}$ & $\to$ & $\mathrm{H_{3}O^{+}}+\mathrm{H}$ & $3.40\times 10^{-9}(T/300)^{-0.5}$\\
	R39 & $\mathrm{H_{2}O}+\mathrm{H_{2}O^{+}}$ & $\to$ & $\mathrm{H_{3}O^{+}}+\mathrm{OH}$ & $2.10\times 10^{-9}(T/300)^{-0.5}$\\
	R40 & $\mathrm{H_{2}O}+\mathrm{H_{3}^{+}}$ & $\to$ & $\mathrm{H_{3}O^{+}}+\mathrm{H_{2}}$ & $5.90\times 10^{-9}(T/300)^{-0.5}$\\
	R41 & $\mathrm{H_{2}O}+\mathrm{OH^{+}}$ & $\to$ & $\mathrm{H_{3}O^{+}}+\mathrm{O}$ & $1.30\times 10^{-9}(T/300)^{-0.5}$\\
	R42 & $\mathrm{H_{2}O}+\mathrm{H}$ & $\to$ & $\mathrm{OH}+\mathrm{H_{2}}$ & $1.59\times 10^{-11}(T/300)^{1.2}\mathrm{exp}(-9600/T)$\\
	R43 & $\mathrm{H_{2}O}+\mathrm{O}$ & $\to$ & $\mathrm{OH}+\mathrm{OH}$ & $1.85\times 10^{-11}(T/300)^{0.95}\mathrm{exp}(-8571/T)$\\	
	R44 & $\mathrm{H}+\mathrm{OH}$ & $\to$ & $\mathrm{O}+\mathrm{H}+\mathrm{H}$ & $6.00\times 10^{-9}\mathrm{exp}(-50900/T)$\\
	R45 & $\mathrm{H}+\mathrm{H_{2}^{+}}$ & $\to$ & $\mathrm{H_{2}}+\mathrm{H^{+}}$ & $6.40\times 10^{-10}$\\
	R46 & $\mathrm{H}+\mathrm{O^{+}}$ & $\to$ & $\mathrm{O}+\mathrm{H^{+}}$ & $5.66\times 10^{-10}(T/300)^{0.36}\mathrm{exp}(8.6/T)$\\
	R47 & $\mathrm{H}+\mathrm{H^{+}}$ & $\to$ & $\mathrm{H_{2}^{+}}$ & $1.15\times 10^{-18}(T/300)^{1.49}\mathrm{exp}(-228/T)$\\
	R48 & $\mathrm{H}+\mathrm{O_{2}}$ & $\to$ & $\mathrm{O}+\mathrm{O}+\mathrm{H}$ & $6.0\times 10^{-9}\mathrm{exp}(-52300/T)$\\
	R49 & $\mathrm{H}+\mathrm{O_{2}}$ & $\to$ & $\mathrm{OH}+\mathrm{O}$ & $2.61\times 10^{-10}\mathrm{exp}(-8156/T)$\\
	R50 & $\mathrm{H}+\mathrm{O}$ & $\to$ & $\mathrm{OH}$ & $9.90\times 10^{-19}(T/300)^{-0.38}$\\
	\hline
	 & & & & Ref. McElroy et al. (2013)\\
	\end{tabular}
\end{table*}	
	
\begin{table*}
	\centering
	\caption{Chemical reactions (continued)}
	\label{tab:A2}
	\begin{tabular}{llcll} 
	\hline
	No. & Reaction &  &   & Reaction rate $\mathrm{cm^{3}\,s^{-1}}$\\
	\hline
	R51 & $\mathrm{H}+\mathrm{OH}$ & $\to$ & $\mathrm{H_{2}O}$ & $5.26\times 10^{-18}(T/300)^{-5.22}\mathrm{exp}(-90/T)$\\
	R52 & $\mathrm{H}+\mathrm{OH}$ & $\to$ & $\mathrm{O}+\mathrm{H_{2}}$ & $6.99\times 10^{-14}(T/300)^{2.8}\mathrm{exp}(-1950/T)$\\	
	R53 & $\mathrm{O}+\mathrm{H^{+}}$ & $\to$ & $\mathrm{O^{+}}+\mathrm{H}$ & $6.86\times 10^{-10}(T/300)^{0.26}\mathrm{exp}(-224.3/T)$\\
	R54 & $\mathrm{O}+\mathrm{H_{2}^{+}}$ & $\to$ & $\mathrm{OH^{+}}+\mathrm{H}$ & $1.50\times 10^{-9}$\\
	R55 & $\mathrm{O}+\mathrm{H_{3}^{+}}$ & $\to$ & $\mathrm{H_{2}O^{+}}+\mathrm{H}$ & $3.42\times 10^{-10}(T/300)^{-0.16}\mathrm{exp}(-1.4/T)$\\
	R56 & $\mathrm{O}+\mathrm{H_{3}^{+}}$ & $\to$ & $\mathrm{OH^{+}}+\mathrm{H_{2}}$ & $7.98\times 10^{-10}(T/300)^{-0.16}\mathrm{exp}(-1.4/T)$\\
	R57 & $\mathrm{O}+\mathrm{H_{2}O^{+}}$ & $\to$ & $\mathrm{O_{2}^{+}}+\mathrm{H_{2}}$ & $4.0\times 10^{-11}$\\
	R58 & $\mathrm{O}+\mathrm{OH^{+}}$ & $\to$ & $\mathrm{O_{2}^{+}}+\mathrm{H}$ & $7.10\times 10^{-10}$\\	
	R59 & $\mathrm{O}+\mathrm{OH}$ & $\to$ & $\mathrm{O_{2}}+\mathrm{H}$ & $3.69\times 10^{-11}(T/300)^{-0.27}\mathrm{exp}(-12.9/T)$\\
	R60 & $\mathrm{O}+\mathrm{O}$ & $\to$ & $\mathrm{O_{2}}$ & $4.9\times 10^{-20}(T/300)^{1.58}$\\
	R61 & $\mathrm{O(^{1}D)}$ & $\to$ & $\mathrm{O}$ & $1.0\times 10^{-4}$\\
	R62 & $\mathrm{O(^{1}D)}+\mathrm{H_{2}O}$ & $\to$ & $\mathrm{OH}+\mathrm{OH}$ & $2.2\times 10^{-9}$\\
	R63 & $\mathrm{O(^{1}D)}+\mathrm{H_{2}}$ & $\to$ & $\mathrm{H}+\mathrm{OH}$ & $1.0\times 10^{-9}$\\
	R64 & $\mathrm{O(^{1}D)}+\mathrm{O_{2}}$ & $\to$ & $\mathrm{O_{2}}+\mathrm{O}$ & $3.2\times 10^{-11}\mathrm{exp}(70/T)$\\
	R65 & $\mathrm{O(^{1}D)}+\mathrm{O}$ & $\to$ & $\mathrm{O}+\mathrm{O}$ & $8.0\times 10^{-12}$\\		
	R66 & $\mathrm{OH}+\mathrm{H^{+}}$ & $\to$ & $\mathrm{OH^{+}}+\mathrm{H}$ & $2.1\times 10^{-9}(T/300)^{-0.5}$\\
	R67 & $\mathrm{OH}+\mathrm{H_{2}^{+}}$ & $\to$ & $\mathrm{OH^{+}}+\mathrm{H_{2}}$ & $7.6\times 10^{-10}(T/300)^{-0.5}$\\
	R68 & $\mathrm{OH}+\mathrm{O^{+}}$ & $\to$ & $\mathrm{OH^{+}}+\mathrm{O}$ & $3.6\times 10^{-10}(T/300)^{-0.5}$\\
	R69 & $\mathrm{OH}+\mathrm{H_{2}^{+}}$ & $\to$ & $\mathrm{H_{2}O^{+}}+\mathrm{H}$ & $7.6\times 10^{-10}(T/300)^{-0.5}$\\
	R70 & $\mathrm{OH}+\mathrm{H_{3}^{+}}$ & $\to$ & $\mathrm{H_{2}O^{+}}+\mathrm{H_{2}}$ & $1.3\times 10^{-9}(T/300)^{-0.5}$\\
	R71 & $\mathrm{OH}+\mathrm{O^{+}}$ & $\to$ & $\mathrm{O_{2}^{+}}+\mathrm{H}$ & $3.6\times 10^{-10}(T/300)^{-0.5}$\\
	R72 & $\mathrm{OH}+\mathrm{OH^{+}}$ & $\to$ & $\mathrm{H_{2}O^{+}}+\mathrm{O}$ & $7.0\times 10^{-10}(T/300)^{-0.5}$\\
	R73 & $\mathrm{OH}+\mathrm{H_{2}O^{+}}$ & $\to$ & $\mathrm{H_{3}O^{+}}+\mathrm{O}$ & $6.9\times 10^{-10}(T/300)^{-0.5}$\\
	R74 & $\mathrm{OH}+\mathrm{OH}$ & $\to$ & $\mathrm{H_{2}O}+\mathrm{O}$ & $1.65\times 10^{-12}(T/300)^{1.14}\mathrm{exp}(-50/T)$\\	
	R75 & $\mathrm{O_{2}}+\mathrm{H^{+}}$ & $\to$ & $\mathrm{O_{2}^{+}}+\mathrm{H}$ & $2.0\times 10^{-9}$\\
	R76 & $\mathrm{O_{2}}+\mathrm{H_{2}^{+}}$ & $\to$ & $\mathrm{O_{2}^{+}}+\mathrm{H_{2}}$ & $8.0\times 10^{-10}$\\
	R77 & $\mathrm{O_{2}}+\mathrm{H_{2}O^{+}}$ & $\to$ & $\mathrm{O_{2}^{+}}+\mathrm{H_{2}O}$ & $4.6\times 10^{-10}$\\
	R78 & $\mathrm{O_{2}}+\mathrm{O^{+}}$ & $\to$ & $\mathrm{O_{2}^{+}}+\mathrm{O}$ & $1.9\times 10^{-11}$\\
	R79 & $\mathrm{O_{2}}+\mathrm{OH^{+}}$ & $\to$ & $\mathrm{O_{2}^{+}}+\mathrm{OH}$ & $5.9\times 10^{-10}$\\
	R80 & $\mathrm{H^{+}}+\mathrm{e}$ & $\to$ & $\mathrm{H}+\mathrm{PHOTON}$ & $3.5\times 10^{-12}(T/300)^{-0.75}$\\
	R81 & $\mathrm{H_{2}^{+}}+\mathrm{e}$ & $\to$ & $\mathrm{H}+\mathrm{H}$ & $1.6\times 10^{-8}(T/300)^{-0.43}$\\
	R82 & $\mathrm{H_{3}^{+}}+\mathrm{e}$ & $\to$ & $\mathrm{H_{2}}+\mathrm{H}$ & $2.34\times 10^{-8}(T/300)^{-0.52}$\\
	R83 & $\mathrm{H_{3}^{+}}+\mathrm{e}$ & $\to$ & $\mathrm{H}+\mathrm{H}+\mathrm{H}$ & $4.36\times 10^{-8}(T/300)^{-0.52}$\\
	R84 & $\mathrm{O^{+}}+\mathrm{e}$ & $\to$ & $\mathrm{O}+\mathrm{PHOTON}$ & $3.24\times 10^{-12}(T/300)^{-0.66}$\\
	R85 & $\mathrm{O_{2}^{+}}+\mathrm{e}$ & $\to$ & $\mathrm{O}+\mathrm{O}$ & $1.95\times 10^{-7}(T/300)^{-0.7}$\\
	R86 & $\mathrm{OH^{+}}+\mathrm{e}$ & $\to$ & $\mathrm{O}+\mathrm{H}$ & $3.75\times 10^{-8}(T/300)^{-0.5}$\\	
	R87 & $\mathrm{H_{2}O^{+}}+\mathrm{e}$ & $\to$ & $\mathrm{O}+\mathrm{H_{2}}$ & $3.9\times 10^{-8}(T/300)^{-0.5}$\\
	R88 & $\mathrm{H_{2}O^{+}}+\mathrm{e}$ & $\to$ & $\mathrm{O}+\mathrm{H}+\mathrm{H}$ & $3.05\times 10^{-7}(T/300)^{-0.5}$\\
	R89 & $\mathrm{H_{2}O^{+}}+\mathrm{e}$ & $\to$ & $\mathrm{OH}+\mathrm{H}$ & $8.6\times 10^{-8}(T/300)^{-0.5}$\\
	R90 & $\mathrm{H_{3}O^{+}}+\mathrm{e}$ & $\to$ & $\mathrm{H_{2}O}+\mathrm{H}$ & $7.09\times 10^{-8}(T/300)^{-0.5}$\\
	R91 & $\mathrm{H_{3}O^{+}}+\mathrm{e}$ & $\to$ & $\mathrm{O}+\mathrm{H_{2}}+\mathrm{H}$ & $5.60\times 10^{-9}(T/300)^{-0.5}$\\
	R92 & $\mathrm{H_{3}O^{+}}+\mathrm{e}$ & $\to$ & $\mathrm{OH}+\mathrm{H_{2}}$ & $5.37\times 10^{-8}(T/300)^{-0.5}$\\
	R93 & $\mathrm{H_{3}O^{+}}+\mathrm{e}$ & $\to$ & $\mathrm{OH}+\mathrm{H}+\mathrm{H}$ & $3.05\times 10^{-7}(T/300)^{-0.5}$\\	
	\hline
	 & & & & Ref. McElroy et al. (2013)\\
	\end{tabular}
\end{table*}

\section{Water abundance in the upper atmosphere}
\label{sec:water}

\begin{figure}
    \centering
    \includegraphics[width=\linewidth]{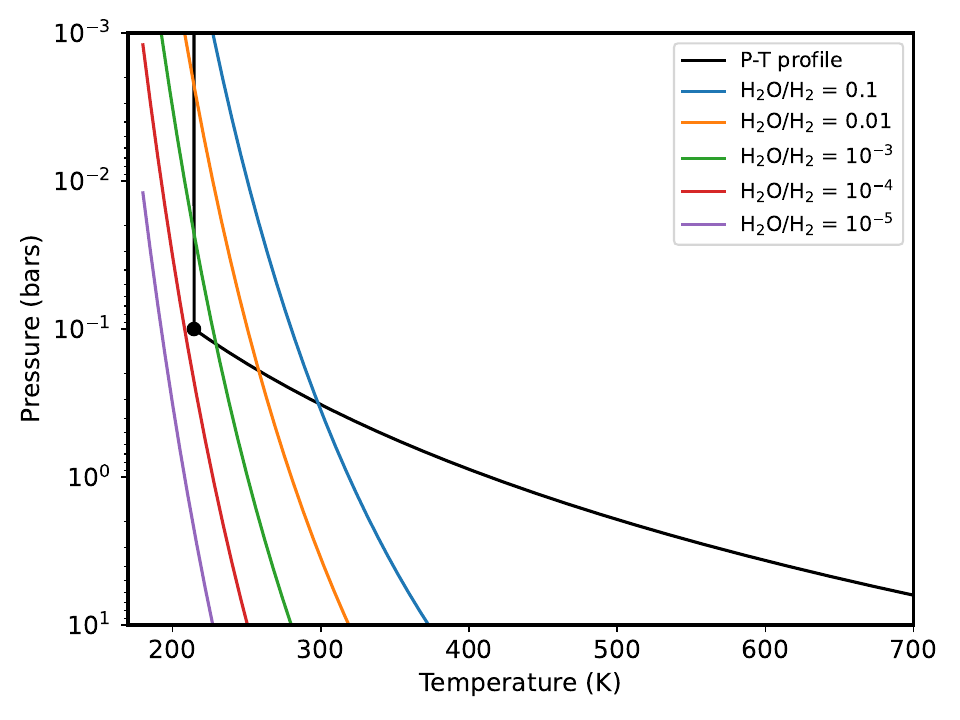}
    \caption{Pressure-temperature profile (black curve) in a simplified $H_2$-rich atmosphere with a radiative-convective boundary at 0.1 bar pressure (black point).  The colored lines represent the vapor pressures for different $H_2$O/H$_2$ mixing ratios, and the lowest vapor pressure along the profile occurs at the R-C boundary (black filled circle).}
    \label{fig:water}
\end{figure}

Even if the lower atmosphere contains plentiful \water\ (as steam), the abundance in the upper atmosphere can be ultimately limited by the saturation vapor pressure where it is lowest relative to the total local pressure.  Figure \ref{fig:water} plots a schematic radiative-convective (R-C) atmosphere where the upper radiative zone is uniformly at the skin temperature $T_{\rm eq}/2^{0.25}$ (here set to the Earth-like value of 255\,K), and the convective zone is an adiabat with $\gamma = 1.4$, appropriate for \htwo.  The R-C boundary (circle) is set at 0.1~bar, a value which is typical for Solar System atmospheres \citep{Robinson2012}.  Curves of the saturation vapor-pressure over ice or liquid \water\ \citep{Bohren1998}, scaled for different mixing ratios, are plotted.  Water vapor in rising atmospheric parcels will condense out and form clouds up to a maximum altitude at the R-C boundary where the vapor pressure is lowest.  The \water/\htwo\ ratio at that boundary and higher altitudes is then set by the saturation vapor pressure curve that intersects the R-C boundary point (in this case, $1.3 \times 10^{-4}$).  Lower or higher $T_{\rm eq}$ results in lower or higher \water/\htwo, accordingly, and this variation describes the curve in Fig. \ref{fig:irrad-water}.  (The curve is not sensitive to the exact location of the R-C boundary).  that the turbopause, above which molecular diffusion dominates over eddy diffusion, is predicted to lie at $\sim10^{-4}$ bar, well above the plot.  

\end{appendix}
\end{document}